\documentclass[10pt, twocolumn, final]{IEEEtran}%

\usepackage[T1]{fontenc}
\usepackage[utf8]{inputenc}
\usepackage{mathtools}
\usepackage{amssymb,mathrsfs}
\usepackage{amsthm}
\usepackage{bm}
\usepackage{scalerel}
\usepackage{nicefrac}
\usepackage{microtype} 
\usepackage[shortlabels]{enumitem}
\usepackage{graphicx}
\usepackage{epstopdf}
\DeclareGraphicsExtensions{.eps,.png,.jpg,.pdf}

\usepackage{url}
\usepackage{colortbl}
\usepackage{booktabs}
\usepackage{multirow}
\usepackage[table,dvipsnames]{xcolor}
\usepackage[normalem]{ulem}
\usepackage{xparse}
\usepackage{calc}
\usepackage{etoolbox}

\makeatletter
\@ifpackageloaded{natbib}{
	\relax
}{
	\usepackage{cite}
}
\makeatother

\usepackage{array}
\newcolumntype{L}[1]{>{\raggedright\let\newline\\\arraybackslash\hspace{0pt}}m{#1}}
\newcolumntype{C}[1]{>{\centering\let\newline\\\arraybackslash\hspace{0pt}}m{#1}}
\newcolumntype{R}[1]{>{\raggedleft\let\newline\\\arraybackslash\hspace{0pt}}m{#1}}

\makeatletter
\let\MYcaption\@makecaption
\makeatother
\usepackage[font=footnotesize]{subcaption}
\makeatletter
\let\@makecaption\MYcaption
\makeatother

\usepackage{glossaries}
\makeatletter
\sfcode`\.1006

\let\oldgls\gls
\let\oldglspl\glspl

\newcommand\fussy@ifnextchar[3]{%
	\let\reserved@d=#1%
	\def\reserved@a{#2}%
	\def\reserved@b{#3}%
	\futurelet\@let@token\fussy@ifnch}
\def\fussy@ifnch{%
	\ifx\@let@token\reserved@d
		\let\reserved@c\reserved@a
	\else
		\let\reserved@c\reserved@b
	\fi
	\reserved@c}

\renewcommand{\gls}[1]{%
\oldgls{#1}\fussy@ifnextchar.{\@checkperiod}{\@}}
\renewcommand{\glspl}[1]{%
\oldglspl{#1}\fussy@ifnextchar.{\@checkperiod}{\@}}

\newcommand{\@checkperiod}[1]{%
	\ifnum\sfcode`\.=\spacefactor\else#1\fi
}

\robustify{\gls}
\robustify{\glspl}
\makeatother

\newacronym{wrt}{w.r.t.}{with respect to}
\newacronym{RHS}{R.H.S.}{right-hand side}
\newacronym{LHS}{L.H.S.}{left-hand side}
\newacronym{iid}{i.i.d.}{independent and identically distributed}

\usepackage{float}

\ifx\notloadhyperref\undefined
	\ifx\loadbibentry\undefined
		\usepackage[hidelinks,hypertexnames=false]{hyperref}
	\else
		\usepackage{bibentry}
		\makeatletter\let\saved@bibitem\@bibitem\makeatother
		\usepackage[hidelinks,hypertexnames=false]{hyperref}
		\makeatletter\let\@bibitem\saved@bibitem\makeatother
	\fi
\else
	\ifx\loadbibentry\undefined
		\relax
	\else
		\usepackage{bibentry}
	\fi
\fi

\usepackage[capitalize]{cleveref}
\crefname{equation}{}{}
\Crefname{equation}{}{}
\crefname{claim}{claim}{claims}
\crefname{step}{step}{steps}
\crefname{line}{line}{lines}
\crefname{condition}{condition}{conditions}
\crefname{dmath}{}{}
\crefname{dseries}{}{}
\crefname{dgroup}{}{}

\crefname{Problem}{Problem}{Problems}
\crefformat{Problem}{Problem~(#2#1#3)}
\crefrangeformat{Problem}{Problems~(#3#1#4) to~(#5#2#6)}

\crefname{Theorem}{Theorem}{Theorems}
\crefname{Corollary}{Corollary}{Corollaries}
\crefname{Proposition}{Proposition}{Propositions}
\crefname{Lemma}{Lemma}{Lemmas}
\crefname{Definition}{Definition}{Definitions}
\crefname{Example}{Example}{Examples}
\crefname{Assumption}{Assumption}{Assumptions}
\crefname{Remark}{Remark}{Remarks}
\crefname{Rem}{Remark}{Remarks}
\crefname{remarks}{Remarks}{Remarks}
\crefname{Appendix}{Appendix}{Appendices}
\crefname{Supplement}{Supplement}{Supplements}
\crefname{Exercise}{Exercise}{Exercises}
\crefname{Theorem_A}{Theorem}{Theorems}
\crefname{Corollary_A}{Corollary}{Corollaries}
\crefname{Proposition_A}{Proposition}{Propositions}
\crefname{Lemma_A}{Lemma}{Lemmas}
\crefname{Definition_A}{Definition}{Definitions}

\usepackage{crossreftools}
\ifx\notloadhyperref\undefined
\else
	\relax
\fi

\usepackage{algorithm}
\usepackage{algpseudocode}

\ifx\loadbreqn\undefined
	\relax
\else
	\usepackage{breqn}
\fi

\makeatletter
\def\cleartheorem#1{%
    \expandafter\let\csname#1\endcsname\relax
    \expandafter\let\csname c@#1\endcsname\relax
}
\def\clearthms#1{ \@for\tname:=#1\do{\cleartheorem\tname} }
\makeatother

\ifx\renewtheorem\undefined
	\ifx\useTheoremCounter\undefined
		\newtheorem{Theorem}{Theorem}
		\newtheorem{Corollary}{Corollary}
		\newtheorem{Proposition}{Proposition}
		\newtheorem{Lemma}{Lemma}
	\else
		\newtheorem{Theorem}{Theorem}

	\fi

	\newtheorem{Definition}{Definition}
	\newtheorem{Example}{Example}

\fi

\theoremstyle{remark}

\theoremstyle{plain}

\newcommand{\qednew}{\nobreak \ifvmode \relax \else
		\ifdim\lastskip<1.5em \hskip-\lastskip
			\hskip1.5em plus0em minus0.5em \fi \nobreak
		\vrule height0.75em width0.5em depth0.25em\fi}



\NewDocumentCommand{\movedownsub}{e{^_}}{%
	\IfNoValueTF{#1}{%
		\IfNoValueF{#2}{^{}}
	}{%
		^{#1}
	}%
	\IfNoValueF{#2}{_{#2}}
}

\let\latexchi\chi
\RenewDocumentCommand{\chi}{}{\latexchi\movedownsub}

\newcommand{\Real}{\mathbb{R}}

\newcommand{\calA}{\mathcal{A}}

\newcommand{\calF}{\mathcal{F}}

\newcommand{\calH}{\mathcal{H}}

\newcommand{\calL}{\mathcal{L}}
\newcommand{\calM}{\mathcal{M}}

\newcommand{\calS}{\mathcal{S}}
\newcommand{\calT}{\mathcal{T}}

\newcommand{\calX}{\mathcal{X}}
\newcommand{\calY}{\mathcal{Y}}
\newcommand{\calZ}{\mathcal{Z}}

\newcommand{\bB}{\mathbf{B}}

\newcommand{\bF}{\mathbf{F}}

\newcommand{\bG}{\mathbf{G}}

\newcommand{\bH}{\mathbf{H}}

\newcommand{\bI}{\mathbf{I}}

\newcommand{\bL}{\mathbf{L}}

\newcommand{\bM}{\mathbf{M}}

\newcommand{\bN}{\mathbf{N}}

\newcommand{\bP}{\mathbf{P}}

\newcommand{\bU}{\mathbf{U}}

\newcommand{\bX}{\mathbf{X}}

\newcommand{\bY}{\mathbf{Y}}

\newcommand{\bZ}{\mathbf{Z}}

\newcommand{\scA}{\mathscr{A}}

\newcommand{\scD}{\mathscr{D}}

\newcommand{\scL}{\mathscr{L}}

\DeclareSymbolFont{bsfletters}{OT1}{cmss}{bx}{n}
\DeclareSymbolFont{ssfletters}{OT1}{cmss}{m}{n}
\DeclareMathSymbol{\bsfGamma}{0}{bsfletters}{'000}
\DeclareMathSymbol{\ssfGamma}{0}{ssfletters}{'000}
\DeclareMathSymbol{\bsfDelta}{0}{bsfletters}{'001}
\DeclareMathSymbol{\ssfDelta}{0}{ssfletters}{'001}
\DeclareMathSymbol{\bsfTheta}{0}{bsfletters}{'002}
\DeclareMathSymbol{\ssfTheta}{0}{ssfletters}{'002}
\DeclareMathSymbol{\bsfLambda}{0}{bsfletters}{'003}
\DeclareMathSymbol{\ssfLambda}{0}{ssfletters}{'003}
\DeclareMathSymbol{\bsfXi}{0}{bsfletters}{'004}
\DeclareMathSymbol{\ssfXi}{0}{ssfletters}{'004}
\DeclareMathSymbol{\bsfPi}{0}{bsfletters}{'005}
\DeclareMathSymbol{\ssfPi}{0}{ssfletters}{'005}
\DeclareMathSymbol{\bsfSigma}{0}{bsfletters}{'006}
\DeclareMathSymbol{\ssfSigma}{0}{ssfletters}{'006}
\DeclareMathSymbol{\bsfUpsilon}{0}{bsfletters}{'007}
\DeclareMathSymbol{\ssfUpsilon}{0}{ssfletters}{'007}
\DeclareMathSymbol{\bsfPhi}{0}{bsfletters}{'010}
\DeclareMathSymbol{\ssfPhi}{0}{ssfletters}{'010}
\DeclareMathSymbol{\bsfPsi}{0}{bsfletters}{'011}
\DeclareMathSymbol{\ssfPsi}{0}{ssfletters}{'011}
\DeclareMathSymbol{\bsfOmega}{0}{bsfletters}{'012}
\DeclareMathSymbol{\ssfOmega}{0}{ssfletters}{'012}

\makeatletter
\newcommand*\rel@kern[1]{\kern#1\dimexpr\macc@kerna}
\newcommand*\widebar[1]{%
  \begingroup
  \def\mathaccent##1##2{%
    \rel@kern{0.8}%
    \overline{\rel@kern{-0.8}\macc@nucleus\rel@kern{0.2}}%
    \rel@kern{-0.2}%
  }%
  \macc@depth\@ne
  \let\math@bgroup\@empty \let\math@egroup\macc@set@skewchar
  \mathsurround\z@ \frozen@everymath{\mathgroup\macc@group\relax}%
  \macc@set@skewchar\relax
  \let\mathaccentV\macc@nested@a
  \macc@nested@a\relax111{#1}%
  \endgroup
}
\makeatother

\newcommand{\ifbcdot}[1]{\ifblank{#1}{\cdot}{#1}}

\DeclarePairedDelimiterX\abs[1]{\lvert}{\rvert}{\ifbcdot{#1}}
\DeclarePairedDelimiterX\parens[1]{(}{)}{\ifbcdot{#1}}
\DeclarePairedDelimiterX\brk[1]{[}{]}{\ifbcdot{#1}}
\DeclarePairedDelimiterX\braces[1]{\{}{\}}{\ifbcdot{#1}}
\DeclarePairedDelimiterX\angles[1]{\langle}{\rangle}{\ifblank{#1}{\cdot,\cdot}{#1}}
\DeclarePairedDelimiterX\ip[2]{\langle}{\rangle}{\ifbcdot{#1},\ifbcdot{#2}}
\DeclarePairedDelimiterX\norm[1]{\lVert}{\rVert}{\ifbcdot{#1}}
\DeclarePairedDelimiterX\ceil[1]{\lceil}{\rceil}{\ifbcdot{#1}}
\DeclarePairedDelimiterX\floor[1]{\lfloor}{\rfloor}{\ifbcdot{#1}}

\DeclareFontFamily{U}{matha}{\hyphenchar\font45}
\DeclareFontShape{U}{matha}{m}{n}{
      <5> <6> <7> <8> <9> <10> gen * matha
      <10.95> matha10 <12> <14.4> <17.28> <20.74> <24.88> matha12
      }{}
\DeclareSymbolFont{matha}{U}{matha}{m}{n}
\DeclareFontSubstitution{U}{matha}{m}{n}

\DeclareFontFamily{U}{mathx}{\hyphenchar\font45}
\DeclareFontShape{U}{mathx}{m}{n}{
      <5> <6> <7> <8> <9> <10>
      <10.95> <12> <14.4> <17.28> <20.74> <24.88>
      mathx10
      }{}
\DeclareSymbolFont{mathx}{U}{mathx}{m}{n}
\DeclareFontSubstitution{U}{mathx}{m}{n}

\DeclareMathDelimiter{\vvvert}{0}{matha}{"7E}{mathx}{"17}
\DeclarePairedDelimiterX\vertiii[1]{\vvvert}{\vvvert}{\ifbcdot{#1}}

\DeclarePairedDelimiterXPP\trace[1]{\operatorname{Tr}}{(}{)}{}{\ifbcdot{#1}} 
\DeclarePairedDelimiterXPP\col[1]{\operatorname{col}}{\{}{\}}{}{\ifbcdot{#1}} 
\DeclarePairedDelimiterXPP\row[1]{\operatorname{row}}{\{}{\}}{}{\ifbcdot{#1}} 
\DeclarePairedDelimiterXPP\erf[1]{\operatorname{erf}}{(}{)}{}{\ifbcdot{#1}}
\DeclarePairedDelimiterXPP\erfc[1]{\operatorname{erfc}}{(}{)}{}{\ifbcdot{#1}}
\DeclarePairedDelimiterXPP\KLD[2]{D}{(}{)}{}{\ifbcdot{#1}\, \delimsize\|\, \ifbcdot{#2}} 
\DeclarePairedDelimiterXPP\op[2]{\operatorname{#1}}{(}{)}{}{#2} 

\newcommand{\ud}{\,\mathrm{d}} 

\DeclarePairedDelimiterXPP\indicate[1]{{\bf 1}}{\{}{\}}{}{\ifbcdot{#1}}

\NewDocumentCommand\ofrac{s m}{%
	\IfBooleanTF#1%
	{\dfrac{1}{#2}}%
	{\frac{1}{#2}}%
}
\NewDocumentCommand\ddfrac{s m m}{%
	\IfBooleanTF#1%
	{\dfrac{\mathrm{d} {#2}}{\mathrm{d} {#3}}}%
	{\frac{\mathrm{d} {#2}}{\mathrm{d} {#3}}}%
}
\NewDocumentCommand\ppfrac{s m m}{%
	\IfBooleanTF#1%
	{\dfrac{\partial {#2}}{\partial {#3}}}%
	{\frac{\partial {#2}}{\partial {#3}}}%
}

\providecommand\given{}

\DeclarePairedDelimiterX\Set[2]\{\}{%
\renewcommand\given{\SetSymbol[\delimsize]{#1}}
#2
}
\DeclarePairedDelimiterX\Setc[1]\{\}{%
\renewcommand\given{\SetSymbol{:}}
#1
}

\NewDocumentCommand\set{s o m}{%
	\IfBooleanTF#1%
	{\IfValueTF{#2}{\Set*{#2}{#3}}{\Setc*{#3}}}%
	{\IfValueTF{#2}{\Set{#2}{#3}}{\Setc{#3}}}%
}

\NewDocumentCommand{\evalat}{ s O{\big} m e{_^} }{%
\IfBooleanTF{#1}%
{\left. #3 \right|}{#3#2|}%
\IfValueT{#4}{_{#4}}%
\IfValueT{#5}{^{#5}}%
}

\providecommand\given{}
\DeclarePairedDelimiterXPP\cprob[1]{}(){}{
\renewcommand\given{\nonscript\,\delimsize\vert\allowbreak\nonscript\,\mathopen{}}%
#1%
}
\DeclarePairedDelimiterXPP\cexp[1]{}[]{}{
\renewcommand\given{\nonscript\,\delimsize\vert\allowbreak\nonscript\,\mathopen{}}%
#1%
}

\DeclareDocumentCommand \P { s e{_^} d() g } {%
	\mathbb{P}%
	\IfBooleanTF{#1}%
		{
			\IfValueT{#2}{_{#2}}%
			\IfValueT{#3}{^{#3}}%
			\IfValueTF{#5}{\cprob{#4 \given #5}}{\IfValueT{#4}{\cprob{#4}}}%
		}%
		{
			\IfValueT{#2}{_{#2}}%
			\IfValueT{#3}{^{#3}}%
			\IfValueTF{#5}{\cprob*{#4 \given #5}}{\IfValueT{#4}{\cprob*{#4}}}%
		}%
}

\DeclareDocumentCommand \E { s e{_^} o g } {%
	\mathbb{E}%
	\IfBooleanTF{#1}%
		{
			\IfValueT{#2}{_{#2}}%
			\IfValueT{#3}{^{#3}}%
			\IfValueTF{#5}{\cexp{#4 \given #5}}{\IfValueT{#4}{\cexp{#4}}}%
		}%
		{
			\IfValueT{#2}{_{#2}}%
			\IfValueT{#3}{^{#3}}%
			\IfValueTF{#5}{\cexp*{#4 \given #5}}{\IfValueT{#4}{\cexp*{#4}}}%
		}%
}

\ExplSyntaxOn
\NewDocumentCommand \dist {m o o} {%
\mathrm{#1}\left(%
	\IfValueT{#3}{%
		\tl_if_blank:nTF{ #3 }{\cdot\, \middle|\, }{#3\, \middle|\, }%
	}
	\IfValueT{#2}{#2}%
\right)%
}
\ExplSyntaxOff

\NewDocumentCommand {\cbrace} {t+ D[]{black} D(){\widthof{#5}} m m } {%
	\begingroup%
		\color{#2}
		\IfBooleanTF{#1}{%
			\overbrace{#4}^%
		}{
			\underbrace{#4}_%
		}%
		{\parbox[c]{#3}{\centering\footnotesize{#5}}}%
	\endgroup%
}

\let\oldforall\forall
\renewcommand{\forall}{\oldforall \, }

\let\oldexist\exists
\renewcommand{\exists}{\oldexist \, }

\newcommand{\figref}[1]{Fig.~\ref{#1}}
\graphicspath{{./Figures/}{./figures/}}
\DeclareDocumentCommand{\includeCroppedPdf}{ o O{./Figures/} m }{
	\IfFileExists{#2#3-crop.pdf}{}{%
		\immediate\write18{pdfcrop #2#3.pdf #2#3-crop.pdf}}%
	\includegraphics[#1]{#2#3-crop.pdf}
}

\makeatletter
\newcommand*{\addFileDependency}[1]{
  \typeout{(#1)}
  \@addtofilelist{#1}
  \IfFileExists{#1}{}{\typeout{No file #1.}}
}
\makeatother

\definecolor{gray90}{gray}{0.9}

\ifx\nohighlights\undefined

	\newcommand{\msout}[1]{\text{\color{green} \sout{\ensuremath{#1}}}}
	\newcommand{\del}[1]{{\color{green}\ifmmode \msout{#1}\else\sout{#1}\fi}}
\else

	\newcommand{\msout}[1]{#1}
	\newcommand{\del}[1]{#1}
\fi

\newcommand{\hhide}[1]{}

\ifx\diagnoselabel\undefined
	\relax
\else
	\makeatletter
	\def\@testdef #1#2#3{%
		\def\reserved@a{#3}\expandafter \ifx \csname #1@#2\endcsname
			\reserved@a  \else
			\typeout{^^Jlabel #2 changed:^^J%
				\meaning\reserved@a^^J%
				\expandafter\meaning\csname #1@#2\endcsname^^J}%
			\@tempswatrue \fi}
	\makeatother
\fi

\crefname{question}{question}{questions}

\usepackage{tikz}
\usepackage{pgfplots}
\pgfplotsset{compat=1.5}

\providecommand{\U}[1]{\protect\rule{.1in}{.1in}}

\theoremstyle{definition}

\begin{document}
\date{}
\title{Graph Signal Processing over a Probability Space of Shift Operators}
\author{Feng~Ji, Wee~Peng~Tay,~\IEEEmembership{Senior Member,~IEEE} and Antonio Ortega,~\IEEEmembership{Fellow,~IEEE}%
\thanks{F.\ Ji and W.\ P.\ Tay are with the School of Electrical and Electronic Engineering, Nanyang Technological University, 639798, Singapore (e-mail: jifeng@ntu.edu.sg, wptay@ntu.edu.sg). They are supported by the Singapore Ministry of Education Academic Research Fund Tier 2 grant MOE-T2EP20220-0002, and the National Research Foundation, Singapore, and Infocomm Media Development Authority under its Future Communications Research and Development Programme. A.\ Ortega is with Viterbi School of Engineering, University of Southern California, Los Angeles, CA 90089-2564  (e-mail: aortega@usc.edu). He is supported in part by the U.S. National Science Foundation under grant NSF CCF-2009032.}%
}
\maketitle

\begin{abstract}
Graph signal processing (GSP) uses a shift operator to define a Fourier basis for the set of graph signals. The shift operator is often chosen to capture the graph topology. However, in many applications, the graph topology may be unknown \emph{a priori}, its structure uncertain, or generated randomly from a predefined set for each observation. Each graph topology gives rise to a different shift operator. In this paper, we develop a GSP framework over a probability space of shift operators. We develop the corresponding notions of Fourier transform, MFC filters, and band-pass filters, which subsumes classical GSP theory as the special case where the probability space consists of a single shift operator. We show that an MFC filter under this framework is the expectation of random convolution filters in classical GSP, while the notion of bandlimitedness requires additional wiggle room from being simply a fixed point of a band-pass filter. We develop a mechanism that facilitates mapping from one space of shift operators to another, which allows our framework to be applied to a rich set of scenarios. We demonstrate how the theory can be applied by using both synthetic and real datasets.
\end{abstract}

\section{Introduction}

Since its emergence, the theory and applications of graph signal processing (GSP) have rapidly developed \cite{Shu12, Shu13, San13, San14, Gad14, Don16, Def16, Egi17, Gra18, Ort18, Girault2018, Ji19}. GSP theory is based on the choice of a \emph{graph shift operator} (GSO) or \emph{fundamental graph operator}, which is a preferred linear transformation on the vector space of graph signals. Once such an operator is given, there is a systematic way to develop a framework for signal processing tasks. To highlight a few important elements of GSP, the change of basis \gls{wrt} an eigenbasis of the shift operator defines the \emph{graph Fourier transform} (GFT) \cite{Shu12, Shu13, Ort18}. The coefficients of a graph signal in the new basis are the components in the \emph{frequency domain}. A central theme of GSP theory is the \emph{theory of filtering} \cite{Shu13, Ort18}, which discusses transformation families. \emph{Convolution} is a transformation by a diagonal matrix in the frequency domain. 
\emph{Sampling} \cite{Aga13, Che15, Tsi15, Mar16, Anis2016, JGT20, Tan20, Lua21} refers to observing only a subset of vertices and reconstructing the signal under some assumptions on the type of signals to be considered, e.g., bandlimited signals.  
\begin{figure}[!htb] 
\centering
\includegraphics[width=0.5\columnwidth]{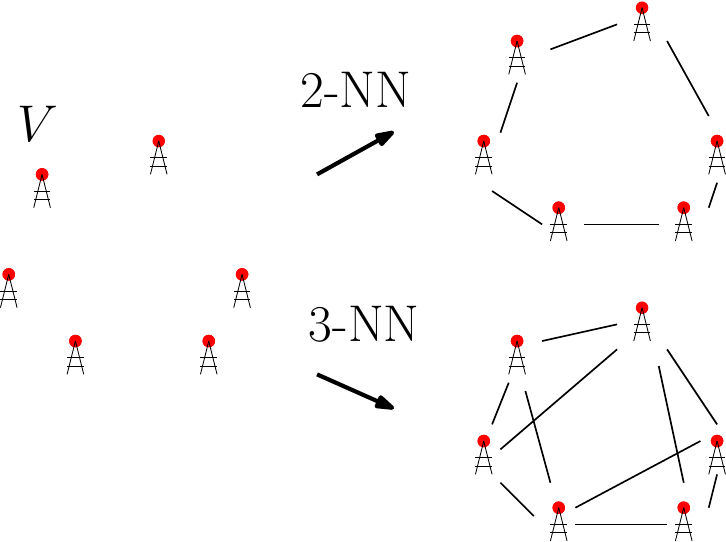}
\caption{Suppose $V$ is a sensor network (left). There are different possible connections among $V$, as shown on the right.} \label{fig:sensors}
\end{figure}

Many techniques have been developed in recent years to learn graphs from data \cite{Mat19, Don19}. 
Roughly speaking, some methods consider features associated with each node, and derive graph edges from the distances between feature vectors, while others learn a graph such that a series of signal examples have some desirable properties, such as smoothness \cite{ortega_2022}. 
Examples of the former type of methods include $k$ nearest neighbors ($k$-NN) and its variations \cite{Alt92, San13}, while the latter include techniques based on smoothness \cite{Don16} or precision matrix (inverse covariance) estimation \cite{Kra08, Mij17, Egi17}. In these methods, constructing a graph that is best suited for some downstream task can be accomplished by choosing some design hyperparameters (e.g., $k$ in $k$-NN as illustrated in \cref{fig:sensors}).

Additionally, when the graph construction is based on several heterogeneous feature vectors, the conventional approach is to construct a single graph by essentially giving different weights to each of the features. A popular approach based on this principle is the bilateral filter and related methods \cite{milanfar2012tour}. In the machine learning community, a series of works (cf.\ \cite{Don17}) is based on the idea of constructing a homogeneous graph by choosing only vertices connected by a specific sequence of edge types (known as a meta-path), before further inference or learning. In the case of information transmission over a network, the effective network topology depends not only on physical connections but also on transmission rates $\zeta$ between members of the network. Certain topology inference methods are sensitive to knowing accurately \emph{a priori} such rates \cite{JiT19}. 

Common to all the above methods is the goal of designing \textit{a single graph}, even though that may entail choosing a specific design parameter (e.g., a hyperparameter to learn a graph from data) or combining heterogeneous features (e.g., by choosing the relative weights for the different components in the features associated with a node). In this paper, we introduce a novel approach that avoids having to choose a single graph by instead working with \textit{multiple graphs simultaneously}. Only a few works have started to look at frameworks that deal with multiple graphs using approaches such as the tensorial method \cite{Zha22}, in slightly different contexts. We achieve the above-stated goal by developing graph signal processing on distributions of graphs. 
Thus, instead of selecting a graph corresponding to a single hyperparameter, it is possible to work simultaneously with several graphs derived from different hyperparameters. Similarly, instead of combining heterogeneous features into a single graph, it is possible to proceed with multiple graphs in parallel, selecting instead the relative weights given to the outputs of each of the graphs for the downstream task. 



In this paper, we consider a probability space of graph shift operators for signals on a finite vertex set $V$. 
The choice of a probability distribution may depend on the specific scenario and goals. This choice may represent our prior belief of which shift operator is more likely, fits the data better, or is more compatible with a downstream inference task. The distribution can be associated with hyperparameters such as those in the earlier examples. We develop a ``distribution version'' of the graph Fourier transform and the associated theory of filtering. Our main contributions are as follows:
\begin{itemize}
    \item We introduce a novel GSP framework for distributions on a sample space of shift operators $\mathcal{X}$. We define the Fourier transform given the distribution. This new framework subsumes classical GSP theory as a special case. 
	\item 
 We develop the concept of a mean fiberwise convolution (MFC) filter. It is not convolutional as in \cite{Pus08} but it is an expectation of conventional convolutional graph filters. If $\mathcal{X}$ is parameterized by a real parameter, then an MFC filter can be characterized as a bi-polynomial filter, which is a polynomial filter whose coefficients are themselves polynomials of the parametrization variable of $\mathcal{X}$.
	\item We develop the notion of bandlimited signals, which in general do not form a vector space. We develop bounds for recovering such signals from a subset of vertex signals.
	\item As filters and observed signals may not always be associated with the same probability space of operators, we develop a mechanism that allows us to map from one probability space of operators to another. 
	\item 
 We demonstrate how to apply our framework with several examples:
 i) signal recovery on heterogeneous graphs, where  different graph operators correspond to different choices of node types, 
 ii) sampling and recovery on a weather station network in which $k$-NN is used to construct the graph and various choices of $k$ are possible, 
 and (iii) anomaly detection on an ElectroCorticoGraphy (ECoG) dataset, 
 where the graph is constructed using signal correlations with different correlation thresholds. 
 In each of these cases, our approach allows us to use a graph operator distribution learned from training data. We outperform the single-graph operator GSP approach, even if the single-graph approach can select the best hyperparameter choice using an exhaustive search.
\end{itemize}

While we focus on the theory of signal processing with a given distribution of operators on a network, learning the distribution of operators is itself an important topic if it is not already given by prior knowledge. To develop the distribution models, we make use of an existing Bayesian approach using Markov chain Monte Carlo (MCMC) \cite{Gue19}, but with novel loss functions based on GSP concepts such as the norm of low-frequency signal components. Although we shall focus on the data-driven approach for acquiring distribution information, it is worth mentioning there are also well-known probabilistic graph models such as the Erd\H{o}s-R\'{e}nyi model that plays important roles in many theoretical studies.

The rest of this paper is organized as follows. We introduce the basic setup and define the Fourier transform in \cref{sec:dis}. In \cref{sec:conv} and \cref{sec:ban}, we discuss various families of filters. We present sampling theory in \cref{sec:ban} in conjunction with band-pass filters. In \cref{sec:ba}, we discuss base changes that deal with switching from one sample space of operators to another. The framework requires knowledge of the distribution of the shift operators. In \cref{sec:lea}, we describe ways to learn a distribution if such a priori knowledge is unavailable. We present several numerical examples in \cref{sec:sim} and conclude in \cref{sec:con}. Proofs of all results are deferred to \cref{sec:pro}. A preliminary version of this work was presented in \cite{Ji21}. In this paper, we include more thorough theoretical discussions and further numerical experiments. In Appendix~\ref{sec:rmk}, we compare our approach with algebraic signal processing.

\emph{Notations:} We use $\circ$ to denote function composition. Let $\Real$ denote the set of real numbers, $\Real_+$ the set of non-negative real numbers, $M_n(\Real)$ be the space of $n\times n$ real matrices, and $[n]$ the 
discrete set $\{1,2,\ldots, n\}$. $\E_{\mu}$ is the expectation operator \gls{wrt} the probability measure $\mu$. 
For a Hilbert space $L^2(\Omega)$, we denote its inner product and norm (and associated operator norm) as $\ip{\cdot}{\cdot}_{L^2(\Omega)}$ and $\norm{\cdot}_{L^2(\Omega)}$, respectively. If the Hilbert space is clear from the context, to avoid clutter, we drop the subscripts and use $\ip{\cdot}{\cdot}$ and $\norm{\cdot}$ respectively. We use calligraphic fonts such as $\mathcal{X},\mathcal{Y}, \mathcal{Z}$ for spaces of operators, while the operators are boldfaced. 

\section{Distribution of shifts and the graph Fourier transform} \label{sec:dis}

In this section, we introduce our framework, the corresponding graph Fourier transform, and its left inverse. 

Let $V$ be the set of vertices of a finite graph $G$, where $|V|=n$. A signal on $V$ is a function $f: V \to \mathbb{R}$, where each signal $f(v)$ associates a real value to a vertex $v\in V$. Denote the Hilbert space of such signals by $L^2(V)$, with $\langle f,f' \rangle = \sum_{v\in V}f(v)f'(v)$. It can be identified with $\mathbb{R}^n$ for a fixed ordering of $V$. 
Suppose $\mathcal{X}$ is a metric space of operators on $L^2(V)$, each of which has eigenvectors that form an orthonormal eigenbasis of $\Real^n$, e.g., each $\bX\in \mathcal{X}$ can be represented as an $n \times n$ symmetric matrix. Let $(\mathcal{X}, \scA, \mu_{\mathcal{X}})$ be a probability space, which may be abbreviated as $(\mathcal{X},\mu_{\mathcal{X}})$ if the $\sigma$-algebra $\scA$ is clear from the context.  
We call $(\mathcal{X},\mu_{\mathcal{X}})$ the \emph{base space}. Let $\mathcal{Y}$ be another metric space and consider the product $\mathcal{X}\times \mathcal{Y}$. Then $\{\bX\}\times \mathcal{Y}$ is called the \emph{fiber} at $\bX$, where $\bX\in \mathcal{X}$. In some examples, for ease of presentation, we may abuse terminology by letting $\mathcal{X}$  be a set of objects (e.g., graphs) instead of operators, where each of these objects is associated with a shift operator. 
Moreover, applying $\bX \in \mathcal{X}$ to a signal $f$ is denoted by $\bX(f)$.

As an example of the above setup, suppose that the underlying graph $G$ is random and is generated by a distribution of graphs on $V$. Each $G$ drawn from the distribution gives a Laplacian $\bL_G$, and the collection of such $\bL_G$ yields the space $\mathcal{X}$. The probability measure $\mu_{\mathcal{X}}$ is thus induced by the distribution generating $G$. A specific case is when the edges among vertices in $V$ are fixed, and the probability distribution $\mu_{\mathcal{X}}$ of the graph Laplacian is induced by a probability distribution on the edge weights. If a training set is available, the probability measure $\mu_{\calX}$ can be learned via a Bayesian framework
(see \cref{sec:lea}).

We next introduce the Fourier transform and its (left) inverse. 
Our definition is simply the set of the classical GFT for each shift operator $\bX\in\calX$. This definition however leads to a nontrivial generalization of convolution and band-pass filters in classical GSP.
\begin{Definition} \label{defn:dbn}
For each operator $\bX \in \mathcal{X}$, let $\lambda_{\bX,i}$ be the $i$-th eigenvalue of the operator $\bX$ (ordered by increasing in absolute value) and $u_{\bX,i}$ be an associated eigenvector with $\set{u_{\bX,i} \given i\in[n]}$ forming an eigenbasis of $\Real^n$. Let $L^2(\mathcal{X}\times [n])$ be the $L^2$ Hilbert space endowed with the product measure $\mu_{\mathcal{X}} \times |\cdot|$ where $|\cdot|$ is the counting measure, and the product $\sigma$-algebra $\scA\times 2^{[n]}$ where $2^{[n]}$ is the power set of $[n]$. Define the Fourier transform \gls{wrt} $(\mathcal{X},\mu_{\mathcal{X}})$,
\begin{align} 
\calF_{\mathcal{X}}: L^2(V) \to L^2(\mathcal{X}\times [n])
\end{align}
by $\calF_{\mathcal{X}}(f) = \hat{f}$ where $\hat{f}(\bX,i) = \ip{f}{u_{\bX,i}}$.
\end{Definition} 

The space $L^2(V)$ is an $n$-dimensional vector space. On the other hand, $L^2(\mathcal{X}\times [n])$ can be infinite-dimensional in general. We remark that for readers familiar with tensor product, $L^2(\mathcal{X}\times [n])$ can also be identified with $L^2(\mathcal{X})\otimes \mathbb{R}^n$, though this interpretation is not used in the sequel. As a consequence of (possibly) infinite dimensionality of $L^2(\mathcal{X}\times [n])$, it is impossible for $\calF_\mathcal{X}$ to be invertible. However, it has a left inverse $\calF^{\dagger}_{\mathcal{X}}: L^2(\mathcal{X}\times [n]) \to L^2(V)$ defined as 
\begin{align} \label{eq:psx}
\calF^{\dagger}_{\mathcal{X}}(g) = \int_{\mathcal{X}} \sum_{i=1}^n g(\bX,i)u_{\bX,i} \ud \mu_{\mathcal{X}}(\bX)
\end{align}
for any $g\in L^2(\mathcal{X}\times [n])$.
Note that $\set{u_{\bX,i}\given (\bX,i)\in \mathcal{X}\times [n]}$ are the kernels of the integration. However, unlike GSP, they are not pair-wise orthogonal for different $\bX$. Intuitively, the Fourier transform should contain spectral information w.r.t.\ all the operators $\bX$ in the family $\mathcal{X}$. 
Therefore, in \cref{defn:dbn}, we compute a family of invertible transformations in parallel as compared with classical GSP, and hence the Hilbert space $L^2(\mathcal{X}\times [n])$ is used as the codomain of the Fourier transform. To ensure recovery of the original signal, we need to re-group the signals in the spectral domain by taking a ``weighted average'' (more precisely, an integral), and this leads to $\calF^{\dagger}_\mathcal{X}$.

Regarding $\calF_\mathcal{X}$ and $\calF^{\dagger}_\mathcal{X}$, the following observations can be verified directly from the definitions and we omit the proof.

\begin{Lemma}\label{lem:FTI}
\begin{enumerate}[(a)]
\item $\calF_{\mathcal{X}}$ and $\calF^{\dagger}_\mathcal{X}$ are both well-defined.
\item (Left inverse) $\calF^{\dagger}_{\mathcal{X}} \circ \calF_\mathcal{X}$ is the identity map on $L^2(V)$.
\item (Parseval's identity) $\norm{f}_{L^2(V)} = \norm{\calF_{\mathcal{X}}(f)}_{L^2(\mathcal{X}\times[n])}$ for each graph signal $f\in L^2(V)$. 
\end{enumerate}
\end{Lemma}

Parseval's identity permits the following equivalent interpretation of our framework. The set of vectors $\set{u_{\bX,i}\given (\bX,i)\in \mathcal{X}\times [n]} \subset L^2(V)$ forms a (continuous) \emph{tight frame} of the finite dimensional vector space $L^2(V)$ (cf.\ \cite{Cas00,Ste09}). With this point of view, the operator $\calF_\mathcal{X}$ can be interpreted as an analysis operator, while $\calF^{\dagger}_\mathcal{X}$ is a synthesis operator. 

We want to further analyze the left inverse $\calF^{\dagger}_{\mathcal{X}}$ for use in subsequent sections. We observe that $\calF^{\dagger}_{\mathcal{X}}$ can be decomposed as follows:
\begin{align}
\calF^{\dagger}_{\mathcal{X}}: L^2(\mathcal{X}\times [n]) \xrightarrow{\alpha_{\mathcal{X}}} L^2(\mathcal{X}\times V) \xrightarrow{\beta_{\mathcal{X}}} L^2(V), \label{eq:invdecomposition}
\end{align}
where here for each $\bX\in \mathcal{X}$, $v\in V$, and $g\in L^2(\mathcal{X}\times [n])$,
\begin{align}
\alpha_{\mathcal{X}}(g)(\bX,v) &= \sum_{i=1}^n g(\bX,i) u_{\bX,i}(v), \label{alphaX}
\intertext{$u_{\bX,i}(v)$ is the $v$-component of the eigenvector $u_{\bX,i}$, and for $q\in L^2(\mathcal{X}\times V)$,} 
\beta_{\mathcal{X}}(q)(v) &= \int_{\mathcal{X}} q(\bX,v) \ud\mu_{\mathcal{X}}(\bX). \label{betaX}
\end{align}

Suppose we let $q(\bX,v)=\alpha_{\mathcal{X}}(g)(\bX,v)$. Then $q(\bX,v)$ is the $v$-component of the inverse GFT \gls{wrt} the operator $\bX$, and further applying $\beta_\mathcal{X}$ to $q$ yields $\beta_{\mathcal{X}}(q)(\cdot) = \E_{\mu_{\mathcal{X}}}[q(\bX,\cdot)]$, where the expectation is over $\bX\in \mathcal{X}$. The map $\alpha_\mathcal{X}$ is invertible with its inverse being the fiberwise GFT for each $\bX\in \mathcal{X}$. 

Note that if $g= \calF_{\mathcal{X}}(f)$ for some $f\in L^2(V)$, then in the above discussion, $q(\bX,v)=f(v)$ for all $\bX\in \mathcal{X}$. Hence, $\beta_{\mathcal{X}}(q)(v)$ gives back $f(v)$. The function $\beta_\mathcal{X}$ and hence the above decomposition of $\calF^{\dagger}_\mathcal{X}$ will only output a signal different from $f$ (the one we started with) when we introduce two filter families in subsequent sections:
\begin{enumerate}[(a)]
\item the \emph{MFC filter family} that ``inserts'' a transformation on $L^2(\mathcal{X}\times [n])$ before applying $\alpha_\mathcal{X}$; and
\item a kind of \emph{base change family} that ``inserts'', between $\alpha_\mathcal{X}$ and $\beta_\mathcal{X}$, a transformation $L^2(\mathcal{X}\times V)\to L^2(\mathcal{Z}\times V)$, with $\mathcal{Z}$ also a probability space.
\end{enumerate}

We end this section with two examples. 
\begin{Example} \label{eg:sme}
\begin{enumerate}[(a)]
\item Suppose $\mu_{\mathcal{X}} = \delta_{\bX_0}$ is a distribution concentrated at a single $\bX_0\in \mathcal{X}$. Then $\calF_\mathcal{X}$ is simply the GFT \gls{wrt} the shift operator $\bX_0$ in classical GSP. 
\item \label{it:cti} 
We consider a special case of \cite[Algorithm~3]{Kim17}: Graph Slicing. Suppose the graph $G$ on the vertex set $V$ is fixed. In addition, we have Laplacians $\bL_0$ and $\bL_1$ associated with two subgraphs $G_1, G_2$ of $G$ such that $G_1$ and $G_2$ have disjoint edge sets and $G=G_1\cup G_2$. We form $\mathcal{X}$ parameterized by the unit interval $T = [0,1]$: for each $t \in [0,1]$, let $\bL_t = (1-t)\bL_0+t\bL_1$ be an element of $\calX$. On $T$, $\scA$ is the Borel $\sigma$-algebra and $\mu_{\calX}$ is a probability measure on $T$. We shall use the following setup in subsequent sections. Let $G$ be a $2$D-lattice (e.g., corresponding to the graph of an image). Then $\bL_0$, and $\bL_1$ are the Laplacians of the subgraph containing horizontal and vertical edges only, respectively. Intuitively, we are thinking of vertical and horizontal edges as having different importance for signal processing purposes. 
\end{enumerate}
\end{Example}

\section{Mean fiberwise convolution filters} \label{sec:conv}

In this section, we introduce the family of mean fiberwise convolution (MFC) filters and some important subfamilies, analogous to convolution filters in classical GSP. We show that an MFC filter is an expectation (in the sense of a Bochner integral) of convolution filters in the classical GSP theory. We then show that under some technical conditions, every MFC filter is a bi-polynomial filter, which is a polynomial filter whose coefficients are themselves polynomials of another parameter. \cite{Jif23} Section IV contains more discussions.

Similar to classical GSP, convolutions are defined using the multiplication of functions on the frequency domain. Given $\Gamma \in L^2(\mathcal{X}\times [n])$, multiplication by $\Gamma$ induces a mapping $L^2(\mathcal{X}\times [n]) \to L^2(\mathcal{X}\times [n])$, which we also denote by $\Gamma$. For any $g\in L^2(\mathcal{X}\times[n])$, we define $\Gamma(g)(\bX,i)=\Gamma(\bX,i)g(\bX,i)$ for all $(\bX,i)\in \mathcal{X}\times[n]$. 

\begin{Definition}\label{def:convolution}
Let $\Gamma \in L^2(\mathcal{X}\times [n])$. A \emph{mean fiberwise convolution (MFC) filter} $\bigstar_{\Gamma}: L^2(V) \to L^2(V)$ is defined by the composition $\calF^{\dagger}_{\mathcal{X}}\circ \Gamma\circ\calF_\mathcal{X}$, i.e., for $f\in L^2(V), v\in V$, 
\begin{align}\label{def:conv}
\bigstar_{\Gamma}(f)(v) = \int_{\mathcal{X}} \sum_{i=1}^n \Gamma(\bX,i)\ip{f}{u_{\bX,i}} u_{\bX,i}(v) \ud\mu_{\mathcal{X}}(\bX).
\end{align} 
\end{Definition}

To further understand the expression, for each $\bX\in \mathcal{X}$, we use $\Gamma_{\bX}$ to denote the function in $L^2([n])$ via the formula 
\begin{align}\label{eq:gg}
\Gamma_{\bX}(i) = \Gamma(\bX,i).
\end{align}
Each $\bigstar_{\Gamma_{\bX}}$ is then a convolution filter on $L^2(V)$ in classical GSP theory. This is called a \emph{fiberwise convolution}, and the reason for the term MFC. This filter is nothing but the composition $\alpha_{\mathcal{X}}\circ \Gamma \circ \calF_\mathcal{X}$ evaluated at $(\bX,\cdot)$ (cf.\ \cref{alphaX}). 

We next show that an MFC filter can be expressed as an \emph{expectation} of fiberwise convolutions. 
\begin{Lemma}\label{lem:Gamma_exp}
For any $\Gamma \in L^2(\mathcal{X}\times [n])$, $\bigstar_{\Gamma} = \E_{\mu_{\mathcal{X}}}[\bigstar_{\Gamma_{\bX}}]$ and it is a bounded linear operator.
\end{Lemma}

Note that the expectation of operators here is in the sense of a Bochner integral \cite{HsiEub:15}. Notation-wise, an expectation of operators remains an operator. 

\begin{Example}\label{eg:wds}
In the following examples, we discuss special cases of MFC filters, to highlight some differences between our framework and classical GSP.
\begin{enumerate}[(a)]
\item \label{it:fas} For any signal $g \in L^2(V)$, its Fourier transform $\hat{g}=\calF_{\mathcal{X}}(g)$ belongs to $L^2(\mathcal{X}\times [n])$. It induces an MFC filter $\bigstar_{\hat{g}} : L^2(V) \to L^2(V)$ that maps $f$ to $h = \calF^{\dagger}_{\mathcal{X}}(\calF_{\mathcal{X}}(g)\cdot\calF_{\mathcal{X}}(f))$, i.e., 
\begin{align*}
h(v) = \int_{\mathcal{X}} \sum_{i=1}^n \ip{g}{u_{\bX,i}}\ip{f}{u_{\bX,i}} u_{\bX,i}(v) \ud\mu_{\mathcal{X}}(\bX),
\end{align*} 
for each $v\in V$. A general MFC filter is associated with an element in $L^2(\mathcal{X}\times [n])$. However, not every MFC filter is obtained from an element of $L^2(V)$ as in this example. On the contrary, in classical GSP where $\mathcal{X}$ is a singleton, every convolution operator corresponds to a graph signal, as described in this example.

\item \label{it:iti} If there is a uniform upper bound on the operator norm of $\bX\in \mathcal{X}$, then $\Lambda: (\bX,i) \mapsto \lambda_{\bX,i}$ belongs to $L^2(\mathcal{X}\times [n])$. From \cref{def:conv}, we obtain
\begin{align*}
\bigstar_{\Lambda}(f) & = \int_{\mathcal{X}} \sum_{i=1}^n \lambda_{\bX,i} \ip{f}{u_{\bX,i}} u_{\bX,i} \ud\mu_{\mathcal{X}} \\
& = \int_{\mathcal{X}} \bX(f) \ud\mu_{\mathcal{X}} = \E_{\mu_{\mathcal{X}}}[\bX](f).
\end{align*}
Consequently, $\bigstar_{\Lambda} = \E_{\mu_{\mathcal{X}}}[\bX]$, the expectation of operators in $\mathcal{X}$. More generally, $\bigstar_{\Lambda^k} = \E_{\mu_{\mathcal{X}}}[\bX^k]$ for positive integers $k$. As $\calF_{\mathcal{X}}\circ \calF^{\dagger}_\mathcal{X}$ is not the identity map, ${\bigstar_{\Lambda^k}} \neq {(\bigstar_{\Lambda})^k}$. This is different from classical GSP where $\mu_{\mathcal{X}}$ is concentrated at a single shift operator $\bX_0$.

More concretely, following \cref{eg:sme}\ref{it:cti}, suppose parameters $\bX$ follows the uniform distribution on $[0,1]$. It is straight forward to compute that $({\bigstar_{\Lambda}})^2 = (\bL_0^2+\bL_1^2 + \bL_0\bL_1+\bL_1\bL_0)/4$ and ${\bigstar_{\Lambda^2}} = (2\bL_0^2+2\bL_1^2+\bL_0\bL_1+\bL_1\bL_0)/6$. In general, not only $({\bigstar_{\Lambda}})^2\neq \bigstar_{\Lambda^2}$ are distinct, they do not have a common eigenbasis, i.e., they are not shift invariant with each other.
\end{enumerate}
\end{Example}

Through \cref{eg:wds}, we see differences between MFC filters and convolution filters in classical GSP. It is also worth pointing out that, unlike the classical theory, composing MFC filters does not necessarily yield an MFC filter (cf.\ \cref{sec:imdb}  below). However, there are also common phenomena between them. In classical GSP, under favorable conditions, a convolution filter is always a polynomial of the shift operator. We next discuss an analogy in our case. 

\begin{Lemma} \label{lem:acf}
If almost surely every $\bX\in \mathcal{X}$ does not have repeated eigenvalues, any MFC filter has the form $\E_{\mu_{\calX}}[R(\bX)]$, where $R$ is a mapping $\mathcal{X} \to M_n(\mathbb{R})$ such that $R(\bX)$ is a degree $n-1$ polynomial in $\bX$. 
\end{Lemma}

Under the conditions of \cref{lem:acf}, for almost surely every $\bX\in \mathcal{X}$, the fiberwise convolution $\bigstar_{\Gamma_{\bX}}$ of $\bigstar_{\Gamma}$ is nothing but a polynomial $R(\bX)= \sum_{0\leq i\leq n-1}a_i(\bX)\bX^i$ of degree at most $n-1$ in $\bX$. On the other hand, for each fixed degree $0\leq i\leq n-1$, we may look at the coefficient $a_i(\bX)$ of the $i$-th monomial for each $R(\bX)$, which gives rise to a function on $\bX\in \mathcal{X}$. Motivated by the above discussions, we now consider the following subspace of MFC filters, called \emph{bi-polynomial filters}. Each bi-polynomial filter is an MFC filter, and we are also interested in conditions that ensure an MFC filter is bi-polynomial.

\begin{Definition} \label{defn:sxi}
Suppose $\mathcal{X}$ is parametrized by $T\subset \mathbb{R}$ via a homeomorphism $t \in T\mapsto \bX_t\in \mathcal{X}$. Let the measure induced by $\mu_{\mathcal{X}}$ on $T$ be $\mu_T$. An MFC filter $\bigstar_{\Gamma}$ with $\Gamma \in L^2(\mathcal{X}\times [n])$ is called \emph{a bi-polynomial filter} on $\mathcal{X}$ if for each $\bX_t\in \mathcal{X}$, there are polynomials $a_i(t)$, $0\leq i \leq k$, with degrees (in $t$) bounded by $d$ such that $\bigstar_{\Gamma_{\bX_t}} = \sum_{0\leq i\leq k} a_i(t) \bX_t^i$. We say that its bi-degree is bounded by $(d,k)$. 
\end{Definition}

For the rest of this section, we assume the existence of such a parameter space $T$ as in \cref{defn:sxi}. To give some simple examples, in the $k$-NN construction, the parameter space $T$ for the space $\mathcal{X}$ of graph shifts associated with different values of $k$ can be chosen as the discrete set $\{1,\ldots, n-1\}$, with $n$ being the number of nodes. In \cref{eg:sme}\ref{it:cti}, the parameter space is $T=[0,1]$ as we are taking convex combinations of two given graph shift operators. By \cref{thm:bipoly} below, every MFC filter is a bi-polynomial filter. If such a filter $\bF_t$ has its bi-degree bounded by $(1,1)$, then it takes the explicit form 
$
\bF_t = (a_0 + a_1t)\big(t\bL_1+(1-t)\bL_0)\big) + (b_0 + b_1t)\bI, t\in [0,1], 
$
where $\bL_0$ and $\bL_1$ are defined in \cref{eg:sme}\ref{it:cti}, $\bI$ is the identity transform and $a_0,a_1,b_0,b_1$ are real coefficients.

\begin{Theorem}\label{thm:bipoly}
Suppose $\calX$ is parametrized by $T\subset\Real$, a finite set or bounded interval. If almost surely every $\bX_t \in \mathcal{X}, t\in T$ has no repeated eigenvalues and has uniformly bounded operator norm, then every MFC filter is a bi-polynomial filter.
\end{Theorem}

In classical GSP, convolution filters being polynomial is a useful feature as it facilitates fast computation, and they are readily learned from estimating the coefficients. Moreover, a distributed implementation is possible with either adjacency or Laplacian matrices. Therefore, it is desirable to have a similar phenomenon in our framework as in \cref{thm:bipoly}. 

\section{Band-pass filters and sampling} \label{sec:ban}

In this section, we develop the notion of band-pass filters, which is a special family of MFC filters. We then introduce the concept of bandlimited signals in our framework and discuss their sampling results.

We start by introducing band-pass filters together with the notion of bandlimited signals. Suppose $\mathcal{Y} \subset \mathcal{X}\times [n]$ is a measurable subset. Recall that the indicator function ${\bf 1}_{\mathcal{Y}}$ on $\mathcal{Y}$ is defined as ${\bf 1}_{\mathcal{Y}}(\bY) = 1$ if $\bY\in \mathcal{Y}$ and $0$ otherwise.

\begin{Definition}\label{def:bandpass}
For a measurable subset $\mathcal{Y} \subset \mathcal{X}\times [n]$, the \emph{band-pass filter} $\bB_{\mathcal{Y}}$ w.r.t.\ $\mathcal{Y}$ is defined as the MFC filter associated with ${\bf 1}_{\mathcal{Y}} \in L^2(\mathcal{X}\times [n])$.\footnote{Note that $\bB_{\calY}=\bigstar_{{\bf 1}_{\calY}}$. To simplify notations, we use $\bB_{\calY}$ here instead.} For $\epsilon\geq 0$, the set of \emph{$(\mathcal{Y},\epsilon)$-bandlimited signals} consists of graph signals $f\in L^2(V)$ such that $\norm{\bB_{\mathcal{Y}}(f)-f}\leq \epsilon$.
\end{Definition}

It is important to note that the filter $\bB_{\mathcal{Y}}$ is not a projection in general as it is the expectation of fiberwise projections (cf.\ \cref{lem:Gamma_exp}). This means that $\bB_{\mathcal{Y}}$ may not even have non-zero fixed points $f= \bB_{\mathcal{Y}}(f)$. Therefore, we are not able to define bandlimited signals as the space of fixed points of a band-pass filter as in classical GSP. As a consequence of \cref{def:bandpass}, the set of $(\mathcal{Y},\epsilon)$-bandlimited signals may not be a vector space. However, if $\mu_{\mathcal{X}}$ concentrates on a single $\bX_0 \in \mathcal{X}$ and $\epsilon = 0$ in \cref{def:bandpass}, we recover the theory of band-pass filters and bandlimited signals in classical GSP. 

Having introduced band-pass filters and bandlimited signals, we now discuss sampling theory that studies the recoverability of bandlimited signals from partial signal observations. To start, we have the following basic observation.

\begin{Lemma} \label{lem:tso}
The set of $(\mathcal{Y},\epsilon)$-bandlimited signals is convex, and it is bounded if $\bB_{\calY}$ does not fix any non-zero signal. Moreover, if $\epsilon>0$, then the signal $f(v) \equiv 0$, for all $v\in V$, is an interior point. 
\end{Lemma}

As a consequence of \cref{lem:tso}, if $\epsilon>0$ and $V'$ is a proper subset of $V$, then the signal values at $V'$ of a $(\mathcal{Y},\epsilon)$-bandlimited signal $f$ do not uniquely determine $f$. To see this, \cref{lem:tso} implies that the intersection of $(\mathcal{Y},\epsilon)$-bandlimited signals and the set of vectors with their $V'$ components fixed is an open subset of the latter. In particular, such an intersection is either empty or contains more than one element. Therefore, for signal recovery from sub-samples in the case $\epsilon>0$, we can only aim for approximations instead of exact recovery. 

\begin{Lemma} \label{lem:ate}
All the eigenvalues of $\bB_{\mathcal{Y}}$ are contained within the closed interval $[0,1]$.
\end{Lemma}

Let $0\leq \lambda_1 \leq \ldots \leq \lambda_n\leq 1$ be the eigenvalues of $\bB_{\mathcal{Y}}$ and $u_1,\ldots, u_n$ be the associated eigenvectors chosen to form an orthonormal eigenbasis, since $\bB_{\mathcal{Y}}$ can be represented as an $n\times n$ symmetric matrix. In general, $\lambda_j$, $1\leq j\leq n$ can be distinct, while they can only be either $0$ or $1$ when $\mu_{\mathcal{X}}$ concentrates on a singleton. 

\begin{Lemma} \label{lem:sfi}
Suppose $f = \sum_{1\leq i\leq n}a_iu_i$ is a $(\mathcal{Y},\epsilon)$-bandlimited signal. Then, for $1\leq j \leq n$ such that $\lambda_j\neq 1$, we have $\sum_{1\leq i\leq j}a_i^2\leq \epsilon^2/(1-\lambda_j)^2$. 
\end{Lemma}

As a consequence, if $\lambda_j$ is close to $0$ for some $1\leq j\leq n$, then the components of a signal spanned by $u_1, \ldots, u_j$ have a small contribution.  Therefore, if one wants to find a sampling subset $V_j$ consisting of $n-j$ vertices, one should choose these vertices based on the components of the vectors $u_{j+1},\ldots, u_n$ as follows: Consider the subspace of signals spanned by $u_{j+1},\ldots, u_n$. Let $\bU_{>j}=[u_{j+1},\ldots, u_n]$ be the matrix with columns $u_{j+1},\ldots, u_n$. Each of its rows corresponds to a node in $V$. A subset $V_j \subset V$ of size $n-j$ is called a \emph{uniqueness set} \cite{Anis2016} if the submatrix of $\bU_{>j}$ consisting of the rows of $\bU_{>j}$, corresponding to vertices in $V_j$, is invertible. Let this submatrix be $\bG_{V_j}$, the \emph{recovery matrix}.  If $f$ belongs to the span of $u_{j+1},\ldots, u_n$ (cf.\ bandlimitedness in classical GSP), then $\bU_{>j}\bG_{V_j}^{-1}(f_{V_j})$ recovers the graph signal $f$ perfectly. Denote the operator norm of $\bG_{V_j}^{-1}$ by $\sigma_{V_j}$.

\begin{Theorem} \label{prop:soo}
Suppose observation of a $(\mathcal{Y},\epsilon)$-bandlimited signal $f\in L^2(V)$ is made at a uniqueness set $V_j$, denoted by $f_{V_j}$. Let $f'$ be the linear combination of $u_{j+1},\ldots,u_n$ with coefficients the entries of $\bG_{V_j}^{-1}(f_{V_j})$, i.e., $\bU_{>j}\bG_{V_j}^{-1}(f_{V_j})$. Then:
\begin{enumerate}[(a)]
\item \label{it:nff} $\norm{f'-f} \leq \epsilon(1+\sigma_{V_j})/(1-\lambda_j)$.
\item $f'$ is $(\mathcal{Y},\epsilon')$-bandlimited with 
\begin{align*}
\epsilon'= \epsilon\parens*{1+2\frac{1+\sigma_{V_j}}{1-\lambda_j}}.
\end{align*}
\end{enumerate}
\end{Theorem}

From the above discussions, we see that $\lambda_j$ is an important quantity that controls how well we can recover a bandlimited signal, with a smaller $\lambda_j$ leading to better recovery. 

We have seen the differences between our theory with the traditional GSP theory, for both the notions of ``band-pass filters'' and ``bandlimited signals''. On the other hand, the above notions in traditional theory are ``limits'' of corresponding notions in this section in an appropriate sense. We make this precise in Appendix~\ref{sec:cbs}.

\section{Base change} \label{sec:ba}

So far, we have been dealing with MFC filters exclusively. \cref{lem:Gamma_exp} summarizes an important observation regarding MFC filters: each is an expectation of a ``random variable of fiberwise convolutions'' on $\mathcal{X}$. On the other hand, once we have such random variables, we may pass them from one probability space to another. This yields the base change filters, which we discuss in this section. We first conceptualize base change in an abstract fashion and then describe explicit formulas with illustrated examples in special cases. We start with the basic setup for base change.

Suppose $(\mathcal{Z},\mu_{\calZ})$ and $(\mathcal{X},\mu_{\mathcal{X}})$ are probability spaces of operators. 
In some applications, it is more natural to define a filter family and its distribution on another space $\mathcal{Z}$ while the graph signal is associated with the base space $\mathcal{X}$, or vice versa. A measurable function 
$
h: \mathcal{Z} \to \mathcal{X}
$
allows pushforward to $\mathcal{X}$ or pullback to $\mathcal{Z}$, on which further signal processing is then performed. 

We motivate the need for a base change framework with the following example.

\begin{Example}\label{eg:sap}
Consider an infection (e.g., a piece of information or disease) propagating from a source $s$ in a graph $G=(V,E)$ following the \emph{SI model} \cite{LuoTayLeng13, LuoTayLen14, LuoTayLen:J16}. Any node receiving the infection is called \emph{infected}, and a node remains infected after its first infection. 
A \emph{propagation path} (\figref{fig:dsp8}) is a tree $\calT$ rooted at $s$, which describes precisely how the infection is passed from $s$ to the infected nodes. In the literature, different types of signals can be associated with an infection. For example, with the observation of infection status, one may construct a signal that is $1$ for each infected node and $0$ otherwise (cf.\ \cref{sec:nis}).

An infection spreading model $\zeta$ specifies the probability of infection across each edge in the graph $G$. 
Let $\calZ$ be a set of adjacency matrices, each corresponding to one of the propagation paths in the set $\calT$. 
The spreading model $\zeta$ induces a measure $\mu_{\calZ}$. Parameters such as infection rate (cf.\ \figref{fig:dsp8}) that determine $\zeta$ can often be estimated from sampled data. However, sample data can sometimes be costly to collect, e.g., in disease spreading, the propagation information is inferred from contact tracing, a time-consuming process. While sample data for a specific model $\zeta$ is available, it may not be practical or feasible to obtain data for a different but related spreading model $\zeta'$. For a concrete example, see \cref{sec:nis} where the model $\zeta'$ differs from $\zeta$ on a subset of edges with different transmission rates. Let $\calX$ be the set of propagation path adjacency matrices under $\zeta'$. Are we able to say anything about a corresponding measure $\mu_{\calX}$ for $\calX$? This will involve a pushforward of the learned measure $\mu_{\calZ}$ to the measurable space $\calX$. 
\begin{figure}[!htb] 
\centering
\includegraphics[width=0.5\columnwidth]{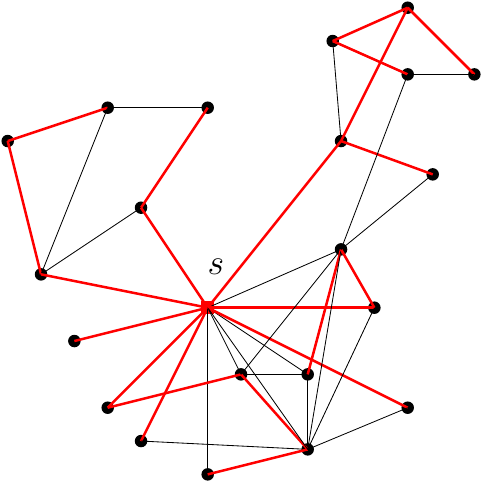}
\caption{In this illustration, the graph $G$ is a part of a social network. Given a source $s$, the red spanning tree is an example of a propagation path. Consider the spreading model whose infection probability follows an exponential distribution $\mathrm{Exp}(\nu)$ with mean $\nu$. A sample propagation path can be generated as follows. We first generate i.i.d.\ ``infection times'' following $\mathrm{Exp}(\nu)$ for edges of $G$. From the source $s$, the associated propagation path is the shortest-path tree (using Dijkstra's algorithm) rooted at $v$.} \label{fig:dsp8}
\end{figure}
\end{Example}

Let $F_{\mathcal{Z}}: \mathcal{Z} \to M_n(\mathbb{R})$ (resp.\ $F_{\mathcal{X}}: \mathcal{X} \to M_n(\mathbb{R})$) denote a mapping of each $\bZ\in \mathcal{Z}$ (resp.\ $\bX\in \mathcal{X}$) to a linear transformation ($n\times n$ matrix). We thus view $F_{\calZ}$ as a family of filters. For example, in the previous sections (cf.\ $\bigstar_{\Gamma_{\bX}}$ before \cref{lem:Gamma_exp}), we have considered MFC filters in which each $F_{\mathcal{X}}(\bX)$ for $\bX \in \mathcal{X}$, is a convolution \gls{wrt} the shift $\bX$. 

We first introduce the pushforward of a measure. In some cases, we may start with a filter family $F_{\calX}$ and would like to have a corresponding family of filters on $\calZ$. This is the pullback of a filter family. These notions are defined formally as follows:
\begin{enumerate}[(a)]
\item \emph{Pushforward of measure}: $h$ and $\mu_{\mathcal{Z}}$ induce a measure $h_*(\mu_{\mathcal{Z}})$ on $\mathcal{X}$, defined as $h_*(\mu_{\mathcal{Z}})(\mathcal{X}') = \mu_{\mathcal{Z}}(h^{-1}(\mathcal{X}'))$ for any measurable subset $\mathcal{X}' \subset \mathcal{X}$.

\item \emph{Pullback of filter family}: $h$ and $F_\mathcal{X}$ induce a family of filters $h^*(F_{\mathcal{X}})$ on $\mathcal{Z}$, defined by $h^*(F_{\mathcal{X}})(\bZ) = F_{\mathcal{X}}(h(\bZ))$.
\end{enumerate}

Similarly, we can define the pullback of measure and pushforward of filters. Here, we need an additional assumption that there is a \emph{fiberwise measure}. More specifically, for each $\bX\in \mathcal{X}$, there is a probability measure $\mu_{h^{-1}(\bX)}$ on the set $h^{-1}(\bX)=\set{\bZ\in \mathcal{Z}\given h(\bZ)=\bX}$. For example, one may choose $\mu_{h^{-1}(\bX)}$ to be the uniform distribution on $h^{-1}(\bX)$. In addition, we assume the technical condition that $\mathcal{Z}$ is locally compact and Hausdorff \cite{Rud87}, which is the case for most of the spaces we are interested in.
\begin{enumerate}[resume*]
\item \label{it:pom} \emph{Pullback of measure}: $h$, $\mu_{\mathcal{X}}$ and $\{\mu_{h^{-1}(\bX)}\}_{\bX\in \mathcal{X}}$ induce a measure $h^*(\mu_{\mathcal{X}})$ on $\mathcal{Z}$ defined by the integration formula 
\begin{align*}
\int_{\mathcal{Z}} g \ud h^*(\mu_{\mathcal{X}}) = \int_{\mathcal{X}} \int_{h^{-1}(\bX)} g \ud\mu_{h^{-1}(\bX)} \ud\mu_{\mathcal{X}}, 
\end{align*} 
where $g$ is any compactly supported continuous function on $\mathcal{Z}$. The measure $h^*(\mu_{\mathcal{X}})$ is uniquely determined by the Riesz–Markov–Kakutani representation theorem \cite{Rud87}.

\item \label{it:pof} \emph{Pushforward of filter family}: $h$, $F_{\mathcal{Z}}$ and $\{\mu_{h^{-1}(\bX)}\}_{\bX\in \mathcal{X}}$ induce a family of filters $h_*(F_{\mathcal{Z}})$ on $\mathcal{X}$ defined by 
\begin{align*} 
h_*(F_{\mathcal{Z}})(\bX) = \int_{h^{-1}(\bX)} F_{\mathcal{Z}}(\bZ) \ud\mu_{h^{-1}(\bX)}(\bZ),\ \bX\in \mathcal{X}. 
\end{align*}
\end{enumerate}



For the rest of this section, we discuss the base change of an MFC filter more concretely and provide explicit formulas. Recall from \cref{eq:invdecomposition} that we have the following decomposition of the identity transform on $\mathcal{X}$: 
\begin{align*} 
\bI = \calF^{\dagger}_{\mathcal{X}}\circ \calF_{\mathcal{X}} = \beta_{\mathcal{X}}\circ \alpha_{\mathcal{X}} \circ \calF_{\mathcal{X}}.
\end{align*} 
We shall insert base changes in this decomposition.

As we have seen, an MFC filter is constructed from $\Gamma \in L^2(\mathcal{X}\times [n])$. The map $h$ pulls it back to a function $h^*(\Gamma) \in L^2(\mathcal{Z}\times [n])$ defined by $h^*(\Gamma)(\bZ,i) = \Gamma(h(\bZ),i)$.  Let $h^\#: L^2(\mathcal{X}\times V) \to L^2(\mathcal{Z}\times V)$ be $h^\#(q)(\bZ,v) = q(h(\bZ),v)$. We have the following associated base change MFCs.

\begin{Definition}
For $\Gamma \in L^2(\mathcal{X}\times [n])$, the filter $\bF_{h^*(\Gamma)}$ defined as the composition 
\begin{align} 
\bF_{h^*(\Gamma)} = \calF^{\dagger}_{\mathcal{Z}}\circ h^*(\Gamma)\circ \calF_{\mathcal{Z}}: L^2(V) \to L^2(V),
\end{align}
is a pullback of the filter $\Gamma$. The filter $\bF_{h^\#,\Gamma}$ is defined as the composition 
\begin{align}
\bF_{h^\#,\Gamma} = \beta_{\mathcal{Z}} \circ h^\# \circ \alpha_{\mathcal{X}}\circ\Gamma\circ \calF_{\mathcal{X}}: L^2(V) \to L^2(V).
\end{align}
\end{Definition}

Recall the eigenbasis $\set{u_{\bX,i} \given i\in[n]}$ in \cref{defn:dbn} for each $\bX\in \mathcal{X}$. Similarly, for each $\bZ\in \mathcal{Z}$, we assume that its eigenvectors form an orthonormal eigenbasis $\set{u_{\bZ,i} \given i\in[n]}$ of $\Real^n$. The filters $\bF_{h^*(\Gamma)}$ and $\bF_{h^\#,\Gamma}$ have the following forms: for $f\in L^2(V)$, 
\begin{align}
\bF_{h^*(\Gamma)}(f) &= \int_{\mathcal{Z}} \sum_{i=1}^n \Gamma(h(\bZ),i)\ip{f}{u_{\bZ,i}} u_{\bZ,i} \ud\mu_{\mathcal{Z}}(\bZ),\\
\bF_{h^\#,\Gamma}(f) &= \int_{\mathcal{Z}} \sum_{i=1}^n \Gamma(h(\bZ),i)\ip{f}{u_{h(\bZ),i}} u_{h(\bZ),i} \ud\mu_{\mathcal{Z}}(\bZ).
\end{align}

If we examine the formulas, we get the intuition that $\bF_{h^\#,\Gamma}$ performs a fiberwise convolution and aggregates according to $\mu_{\mathcal{Z}}$. It can be viewed as a ``re-arrangement'' of ``probability densities''. Effectively, it corresponds to the pushforward of measure. On the other hand, recall that $\Gamma\in L^2(\mathcal{X}\times [n])$ gives rise to a family of filters $\mathcal{X} \to M_n(\mathbb{R})$ by $\bX\mapsto \bigstar_{\Gamma_{\bX}}$. Change of filter family essentially replaces ``$\bX$'' on the subscript by ``$h(\bZ)$''. Hence, as we observe that $\bF_{h^*(\Gamma)}$ ``re-arranges'' the kernel $\Gamma$, it is indeed the pullback of a filter family. Note that $\bF_{h^*(\Gamma)}$ is an MFC filter associated with $\mathcal{Z}$ as in \cref{def:convolution}. However, $\bF_{h^\#,\Gamma}$ may not be an MFC filter, as illustrated in the following example.

\begin{Example} \label{eg:itew}
\begin{enumerate}[(a)]
\item In this example, we consider two cases where either $\mathcal{Z}$ or $\mathcal{X}$ is finite.
\begin{enumerate}[(i)]
\item Suppose $\mathcal{Z}$ is a finite subset of $\mathcal{X}$ and $h: \mathcal{Z} \to \mathcal{X}$ is inclusion, i.e., $h(\bZ)=\bZ$ for all $\bZ\in\calZ$. Then the pushforward measure on $\mathcal{X}$ is a discrete measure supported on $\mathcal{Z} \subset \mathcal{X}$. The pullback of any filter, via $h$, is just the restriction to $\mathcal{Z}$. In this case, $\bF_{h^\#,\Gamma} = \bF_{h^*(\Gamma)}$ performs the following: apply a convolution, with the kernel $\Gamma$ restricted to $\bZ$, at each $\bZ\in \mathcal{Z}$; and then take the expectation according to the discrete measure on $\mathcal{Z}$. The resulting filter is equivalent to the MFC filter on $\mathcal{Z}\times[n] \subset \mathcal{X}\times[n]$.

\item Suppose $\mathcal{Z}$ is parametrized by $T= [0,1]$ with the Lebesgue measure~$\mathfrak{m}$ and $\mathcal{X} = \{\bX_1,\ldots, \bX_k\}$ is parametrized by a finite subset $\{t_1,\ldots, t_k\}$ of $T$. Suppose $T = \bigcup_{1\leq i\leq k} T_i$ has a decomposition into disjoint intervals such that the length $\mathfrak{m}(T_i)>0$ and $t_i \in T_i$ for each $i=1,\ldots,k$. The map $h: \mathcal{Z} \to \mathcal{X}$ is induced by sending the interval $T_i$ to $t_i$, for each $1\leq i\leq k$. The pushforward measure on $\mathcal{X}$ assigns $\mathfrak{m}(T_i)$ to $t_i$, as well as $\bX_i$. If $\Gamma = (\Gamma_{\bX_i} \in \mathbb{R}^n)_{1\leq i\leq k}$ is a function on $\mathcal{X}\times [n]$, then the filter $\bF_{h^\#,\Gamma}$ performs the following: apply a pointwise convolution, with convolution kernel $\Gamma_{\bX_i}$ at each $\bX_i$; and then take the expectation as the weighted sum with weights $\mathfrak{m}(T_i)$ for $1\leq i\leq k$. This is the \emph{coarsening procedure}. Since $h$ is not injective, $\bF_{h^\#,\Gamma}$ is not an MFC filter and hence $\bF_{h^\#,\Gamma}\neq \bF_{h^*(\Gamma)}$. 
\end{enumerate}

\item\label{it:stretch} Recall the setting of \cref{eg:sme}\ref{it:cti}. Suppose $\mathcal{X}, \mathcal{Z}$ are both parametrized by $T_{\mathcal{Z}} = T_{\mathcal{X}}=[0,1]$ equipped with the Lebesgue measure. Let $G$ be a square lattice and let $\bL_0, \bL_1$ be the Laplacians of the subgraphs consisting of horizontal and vertical edges respectively. 
As in \cref{eg:sme}\ref{it:cti} (notice the notations are changed to deal with two spaces), for $x, z\in [0,1]$, we have the matrix $\bL_x = (1-x)\bL_0+x\bL_1$, and similarly $\bL_z = (1-z)\bL_0+z\bL_1$. For $\eta>0$, define $h: T_{\mathcal{Z}} \to T_{\mathcal{X}}$ by the formula $h(z) = z\eta/(1-z+z\eta) \in [0,1]$. The map $h$ is invertible with inverse given by $x \mapsto x/(x+\eta-x\eta)$. It can be verified that if $\bH_x= x\bL_1 + (1-x)(\eta \bL_0)$ as a stretched version of $\bL_x$, then $\bH_x=\eta/(1-z+z\eta) \bL_z $ is a scalar multiple of $\bL_z$ for $x = h(z)$. In particular, $\bH_x$ and $\bL_z$ have the same eigenbasis.

As a consequence, suppose our knowledge of the graph distribution is on the unstretched version $\bL_z$ and the signal is stretched in the horizontal direction by a factor $\eta$. Then to match prior knowledge and the observed signal in a convolution process, one needs to use the filter with base change $\bF_{h^\#,\Gamma}$. An illustration is shown in \figref{fig:bc}
\begin{figure}[!htb] 
\centering
\includegraphics[width=0.6\columnwidth]{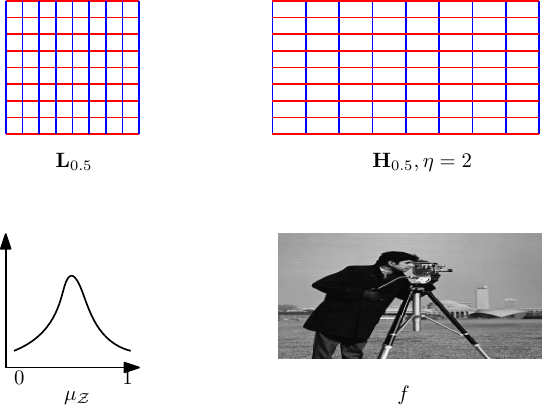}
\caption{The figures show the case $\eta=2$, i.e., the graph is stretched horizontally by a factor of $2$. Suppose $f$, e.g.\ the image below, is stretched and knowledge of the graph distribution is based on unstretched graphs. To correctly perform signal processing, one needs to apply base change.} \label{fig:bc}
\end{figure}	
\end{enumerate}
\end{Example}

\section{Learning the shift distribution} \label{sec:lea}
The framework discussed in this paper relies on knowing a probability space $(\mathcal{X}, \mu_{\mathcal{X}})$ of shift operators. When such information is not directly available, we propose a Bayesian learning framework to estimate $(\mathcal{X},\mu_{\mathcal{X}})$ from sampled data, which essentially involves the MCMC method. For this, we follow largely \cite{Gue19}, which we briefly discuss here for completeness.

Suppose we have a candidate space $\mathcal{X}$ and want to learn a discrete distribution $\mu_{\mathcal{X}}$ that approximates the true distribution. We need the following data: 
\begin{itemize}
\item There is a set of training signals: $\scD = \set{f_i \in L^2(V)\given 1\leq i\leq m}$, possibly with labels $\scL = \set{z_i\in\Real \given 1\leq i\leq m}$. 
\item There is a loss function to be minimized: $\ell: M_n(\mathbb{R})\times L^2(V) \to \mathbb{R}_+$ in the unlabeled case, and $\ell : M_n(\mathbb{R})\times L^2(V)\times \Real \to \Real_+$ in the labeled case.
\item There is a prior measure $\mu_0$ on $\mathcal{X}$. We usually choose the uninformative prior, i.e., the uniform measure.
\end{itemize}

For each $\bX\in \mathcal{X}$, we may now define the empirical risk as
$\theta(\bX) = \frac{1}{m} \sum_{1\leq i\leq m}\ell(\bX,f_i)$
in the unlabeled case, and
$\theta(\bX) = \frac{1}{m} \sum_{1\leq i\leq m}\ell(\bX,f_i,z_i)$
in the labeled case.

Suppose $\mu_0$ has density function $p_0$ (\gls{wrt} some dominating measure like Lebesgue measure). For a fixed parameter $\gamma > 0$, discrete samples to approximate $\mu_{\mathcal{X}}$ are drawn proportional to $\exp\parens*{-\gamma \theta(\cdot)}p_0(\cdot)$, yielding the Gibbs posterior. The main insight from \cite{Gue19} is that if we treat $\ell$ as a prediction loss, then the learned distribution is an approximation of the actual distribution in the following sense: the expected ``prediction'' with the learned distribution has a good average performance. The exact statements are called the PAC-Bayesian inequalities \cite[Section 4]{Gue19}. To generate such samples, one may use the Metropolis-Hastings algorithm \cite{Has70}. 
For each application, it is important to choose an appropriate loss function $\ell$. As $\ell$ depends largely on the explicit situation, we shall describe case-by-case choices in \cref{sec:sim}. 

We next discuss base changes in the unlabeled case, in parallel with \cref{sec:ba}. The labeled case can be treated similarly. As earlier, we assume that there is a measurable function between measure spaces $h: \mathcal{Z} \to \mathcal{X}$. We can pushforward a measure from $\mathcal{Z}$ to $\mathcal{X}$ or pullback a measure from $\mathcal{X}$ to $\mathcal{Z}$ as in \cref{sec:ba}, depending on whether the training data $\scD$ is associated with $\mathcal{X}$ or $\mathcal{Z}$. 

For $\ell$, the pullback is natural to define without additional assumptions. Specifically, $h^*(\ell)$ is defined as $h^*(\ell)(\bZ,f) = \ell(h(\bZ),f)$. It induces pullback of the risk $h^*(\theta)$. As earlier, to define the pushforward $h_*$, we require fiberwise measures $\{\mu_{h^{-1}(\bX)}\}_{\bX\in \mathcal{X}}$, and $h_*(\ell)(\bX,f) = \int_{h^{-1}(\bX)}\ell(\cdot,f)\ud\mu_{h^{-1}}(\bX)$. It induces pushforward of the risk $h_*(\theta)$. Therefore, depending on the situation, the learned $\mu_{\mathcal{X}}$, using the framework of \cite{Gue19} as described above, is proportional to one of the following: 
\begin{itemize}
    \item $\exp(-\gamma \theta(\cdot))p_0(\cdot)$,
    \item $\exp(-\gamma \theta(\cdot))h_*(p_0)(\cdot)$,
    \item $\exp(-\gamma h_*(\theta)(\cdot))p_0(\cdot)$,
    \item $\exp(-\gamma h_*(\theta)(\cdot) \big)h_*(p_0)(\cdot)$.
\end{itemize}   
For $\mu_{\mathcal{Z}}$, we just replace $h_*$ by $h^*$ in the above expressions.

\begin{Example} \label{eg:mni}
We study the MNIST dataset\footnote{http://yann.lecun.com/exdb/mnist/} using the setup of \cref{eg:sme}\ref{it:cti}. This means that we use a $2$D-lattice $G=(V,E)$ to model each image. Let $\bL_0$ and $\bL_1$ be the Laplacians of subgraphs of $G$ consisting of horizontal and vertical edges, respectively. The motivation is that contributions, in terms of characterizing functionality, from horizontal and vertical edges can be different for different digits. For example, for the digit $1$, vertical edges might be more important, while for $0$, both horizontal and vertical edges may play similar roles.

$\mathcal{X}$ is parametrized by the unit interval $[0,1]$; and $t\in [0,1]$ gives rise to $\bL_t = (1-t)\bL_0+t\bL_1$. The size of each image is $28\times 28$. The pixel values can be viewed as a signal $f$ on $V$.  We find a distribution on $\mathcal{X}$ based on sparse encoding of $f$ with each $t\in [0,1]$. More precisely, let $\hat{f}(t,\cdot) : [784] \to \mathbb{R}$ be the Fourier transform of $f$ w.r.t.\ $\bL_t$. The loss function is: 
\begin{align*}
\ell(\bL_t,f)^2 = \frac{\sum_{400\leq i\leq 784} |\hat{f}(t,i)|^2}{\norm{f}^2}.
\end{align*}

For each digit $j=0,1,\ldots, 9$, the empirical distribution $\mu_{j,\mathcal{X}}$ of $\mathcal{X}$ is shown in \figref{fig:dsp5}. It is interesting to observe that for several digits, e.g., digit $1$, there is an obvious shift of the distribution away from the center $t=0.5$. 

\begin{figure}[!htb] 
\centering
\includegraphics[width=0.8\columnwidth,trim=0 7.6cm 0 7.5cm,clip]{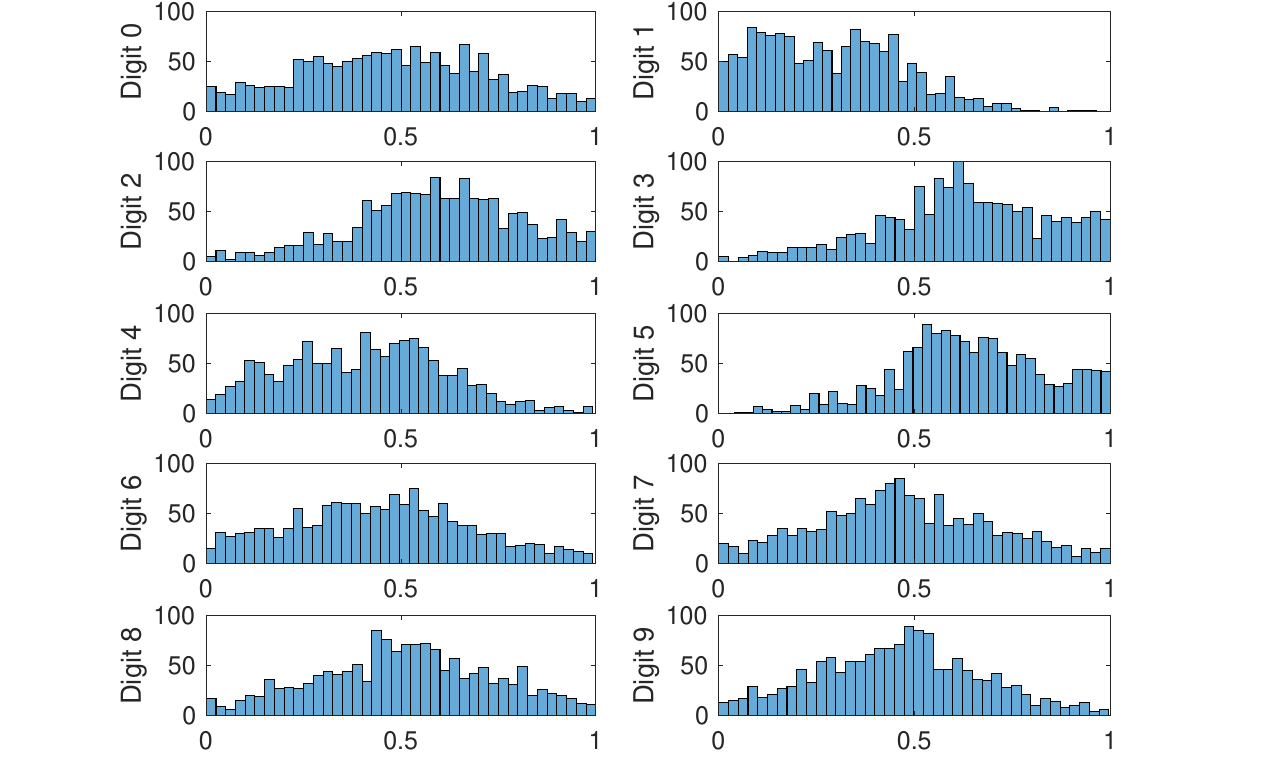}
\caption{Histogram of samples for each digit, with $t$ as the $x$-axis and the number of samples as the $y$-axis for each subplot.} \label{fig:dsp5}
\end{figure}	

For an $f$, we add i.i.d.\ Gaussian noise to each pixel, and the resulting signal is denoted by $f'$. On each of the distributions $\mu_{j,\mathcal{X}}$, $j=0,\ldots,9$, we design a simple MFC filter $\bigstar_{\Gamma_j}$ induced by $\Gamma_j(t,i)=1$ if $i\leq 400, t\in [0,1]$ and $\Gamma_j(t,i)=0.1$ if $i> 400, t\in [0,1]$ (cf.\ \cite{San14}). We apply $\bigstar_{\Gamma_j}$ to the noisy image $f'$ and examples are shown in \figref{fig:dsp6}. Alongside, we also show the image $f'$ and the image signal obtained by applying an ordinary convolution filter $\bigstar_{\Gamma_G}$ (constructed similarly as above) with $\bL_G = 2\bL_{0.5}$. We see that in general $\bigstar_{\Gamma_j}$ produces arguably sharper images of the digits, with the contrast between a digit and its surrounding region higher.

\begin{figure*}
  \hspace*{-2cm} 
  \includegraphics[width=2.3\columnwidth,trim=0 8.6cm 0 8.5cm,clip]{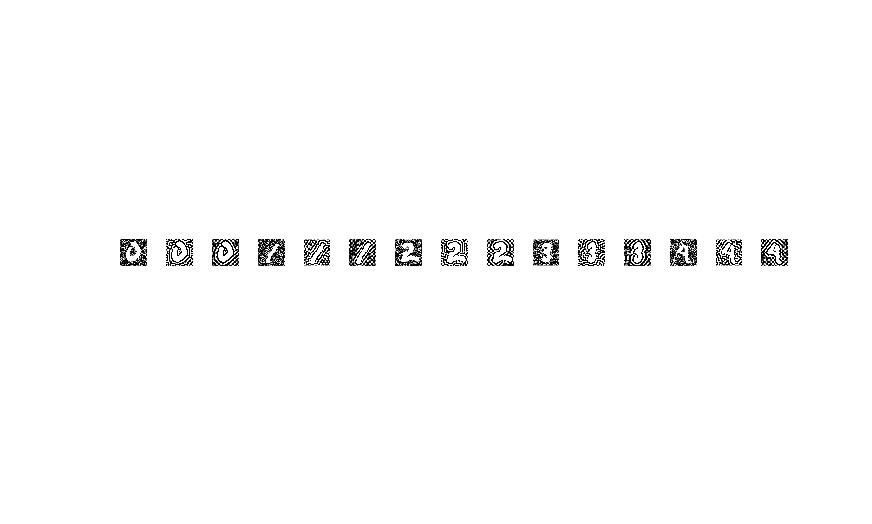}
  \hspace*{-2cm} 
  \includegraphics[width=2.3\columnwidth,trim=0 8.6cm 0 8.5cm,clip]{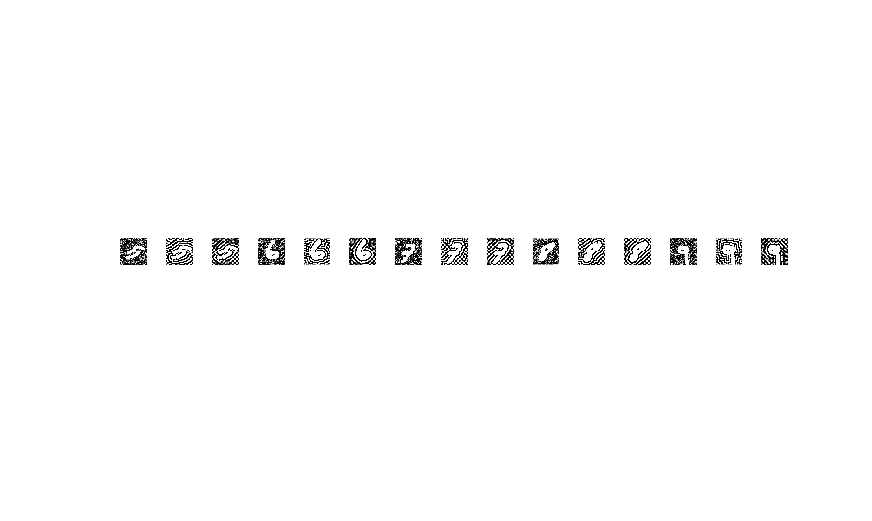}
  \caption{For each digit, there are three images, corresponding to the noisy image $f'$, the image processed with $\bigstar_{\Gamma_G}$ and the image processed with $\bigstar_{\Gamma_j}$, respectively from left to right.} \label{fig:dsp6}
\end{figure*}
\end{Example}

\section{Numerical results} \label{sec:sim}

In this section, we present simulation results. We first present an example where the graph construction can be based on different attributes. We then consider a sampling and recovery problem on a weather station dataset where the underlying graph can be constructed using a $k$-NN approach, and anomaly detection on an ECoG dataset where graphs are constructed by thresholding pairwise node correlations. Finally, we present a network infection spreading example to illustrate the use of base changes. The main purpose is to showcase how the framework proposed in this paper can be applied. In each example, aside from its specific purpose, we emphasize how the space of operators arises and how information about the distribution is acquired.

\subsection{IMDB dataset: heterogeneous graphs} \label{sec:imdb}
In this subsection, we demonstrate an application of MFC filters discussed in \cref{sec:conv}. The dataset considered is the IMDB dataset\footnote{\url{https://www.imdb.com/interfaces/}} used for the study of heterogeneous graphs \cite{Yun19, Lee22}, where nodes and edges have different types. In our case, we associate movies in the IMDB dataset with nodes $V$ of a graph. We form two graphs: an actor graph $G_a$ where two nodes are connected by an edge if they share a common actor (or actress), and a director graph $G_d$ where two nodes are connected if they share a common director. The graph $G_a$ is denser as each movie has multiple actors. Movies are categorized into a few classes, resulting in a signal $f$ on $V$ consisting of integer labels. In practice, there can be data corruption even in the storage centers of large tech corporations, due to reasons such as temperature variance, aging facilities, and mismanagement. Errors may occur without leaving any trace in system logs \cite{Kwa22}. The node labeling process itself could also be noisy, e.g., if the dataset is labeled using crowdsourcing \cite{Karger2013,KanTay:J19}. In our setup, it can be useful to perform correction without knowing the exact node identities where corruption occurs. To simulate, we assume $f$ is corrupted by (integer) noise. More specifically, we add independent additive white Gaussian noise to each entry of $f$. Each entry in the resulting signal is then rounded to the nearest integer. Let the final corrupted signal be $f_c$. In our experiments, we add different amounts of noise to vary the SNR as $-5$~dB and $-1$~dB, so that $f_c$ has a significant number of wrong labels.

Let $\bL_{G_a}, \bL_{G_d}$ be the Laplacians of $G_a$ and $G_d$, respectively.
We apply convolution (resp.\ MFC) filters to $f_c$ to recover $f$. We consider three different frameworks below, all of which use a common filter construction approach. Let $r_1, r_2 \geq 0$ be scaling factors and $c \in [n]$ be a cutoff threshold. These hyperparameters are denoted as $\omega = \{r_1,r_2,c\}$. The convolution (resp.\ MFC) filter we use is to apply the mask $g_{\omega}$ in the graph frequency domain, where $g_{\omega}(i) = r_1$ if $i\leq c$ and $g_{\omega}(i) = r_2$ if $i>c$. The hyperparameters $\omega$ are tuned using $30$ samples to achieve the optimal performance, and hence they can be different for different frameworks.
The frameworks we test are:
\begin{enumerate}[F1]
\item \label{it:cgw} Classical GSP with $\bL_{G_a}$ or $\bL_{G_d}$ as the shift operator. 
\item \label{it:cgwt} Classical GSP with the graph shift operator being a mixture of $\bL_{G_a}$ and $\bL_{G_d}$, i.e., we choose an operator from $\set{\bL_t= t\bL_{G_a} + (1-t)\bL_{G_d}\given t\in (0,1)}$. We apply the same convolution filter given by the frequency mask $g_{\omega}$ as above. In our experiment, we perform an exhaustive search over $19$ uniformly spaced $t$s in $(0,1)$ and show the result for the operator $\bL_{t^*}$ with the best performance.
\item Proposed framework with $\calX = \{\bL_{G_a}, \bL_{G_d}\}$ and $\mu_{\calX}$ uniform over $\calX$. We use the MFC filter $\bigstar_{\Gamma}$ where $\Gamma(\bL_G,\cdot) = g_{\omega_G}(\cdot), G \in \set{G_a, G_d}$. 
\end{enumerate}

\begin{figure}[!htb] 
\centering
\includegraphics[width=0.70\columnwidth, trim=0.1cm 0cm 0cm 0cm,clip]{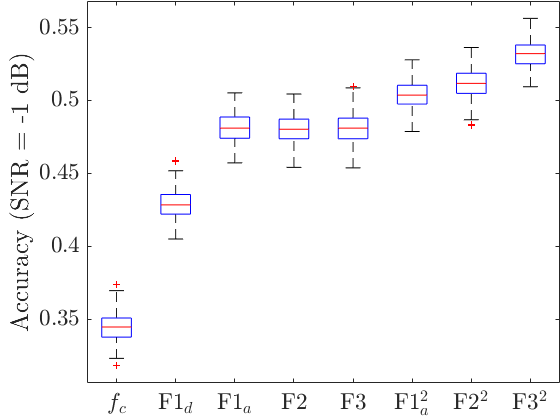}
\includegraphics[width=0.70\columnwidth, trim=0.1cm 0cm 0cm 0cm,clip]{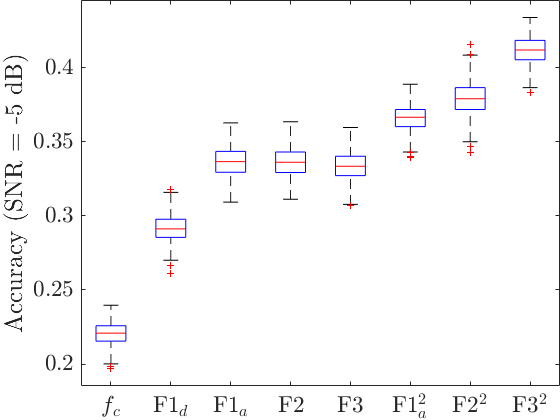}
\caption{Boxplots of recovery accuracy (for IMDB). F1$_d^2$ is not shown as it is almost identical to F1$_d$.}\label{fig:1}
\end{figure}

We also consider their compositional versions F1$^2$, F2$^2$, and F3$^2$, where we compose two filters (with possibly different parameters) of their respective types in F1, F2, and F3. We write F1$_a$ and F1$_d$ for F1 that uses $\bL_{G_a}$ and $\bL_{G_d}$, respectively. We show the label accuracy of the recovered signal in \cref{fig:1}. Each plot uses $300$ $f_c$'s with the same noise level. 

We notice that F1$_a$, F2, and F3 have similar performance without composition. This may be due to the fact that $G_d$ is sparse and it does not make a significant contribution if we linearly combine $\bL_{G_a}$ and $\bL_{G_d}$ or take an expectation of their bandlimited filters. The sparsity of $G_d$ also accounts for the low accuracy of F1$_d$.

If we compose two filters, F3$^2$ has a better performance. By classical GSP, both F1 (resp.\ F2) and F1$^2$ (resp.\ F2$^2$)  consider a single variable polynomial of a fixed GSO $\bL_{G_a}$ or $\bL_{G_d}$ (resp.\ $\bL_{t^*}$). On the other hand, F3$^2$ involves a two-variable polynomial on two distinct GSOs. It contains terms taking the form of products of $\bL_{G_d}$ and $\bL_{G_a}$, which are missing in F3. Therefore, F3$^2$ may capture additional useful interactions between the two GSOs. A similar consideration (to F3$^2$) can be found in graph transformer networks (GTN) \cite{Yun19}, where each transformer layer involves a component similar to that of F3 (cf.\ \cite[eq.\ (4) and eq.\ (8)]{Yun19}) and  multiple layers are stacked, which is analogous to composing filters.

Though models such as GTN in \cite{Yun19} can be more powerful, bandlimited filters are simple and interpretable. Moreover, they can be used to demonstrate the key differences between the proposed framework and classical GSP.

\subsection{Weather station network: sampling and recovery} \label{sec:weather}

In this experiment, we study sampling and recovery (cf.\ \cref{sec:ban}). The network is a real weather station network in the United States with $n=194$ nodes.\footnote{http://www.ncdc.noaa.gov/data-access/} Sampling and recovery techniques can be useful in cases of failure of or inaccessibility to certain stations, due to reasons such as system malfunction and extreme weather conditions. It can also be helpful to reduce sensor operation and data storage. Though geographic distances between pairs of stations are available, there is no explicit graph connecting the stations. The signals are based on daily temperature readings over 2013. By preliminary inspection, we notice that the signals are smooth. We want to estimate temperature readings over the entire network based on those sampled at $5$ stations. 

\begin{figure}[!htb] 
\centering
\includegraphics[width=0.65\columnwidth,height=.16\textheight,trim=0.5cm 6.5cm 0 6.5cm, clip]{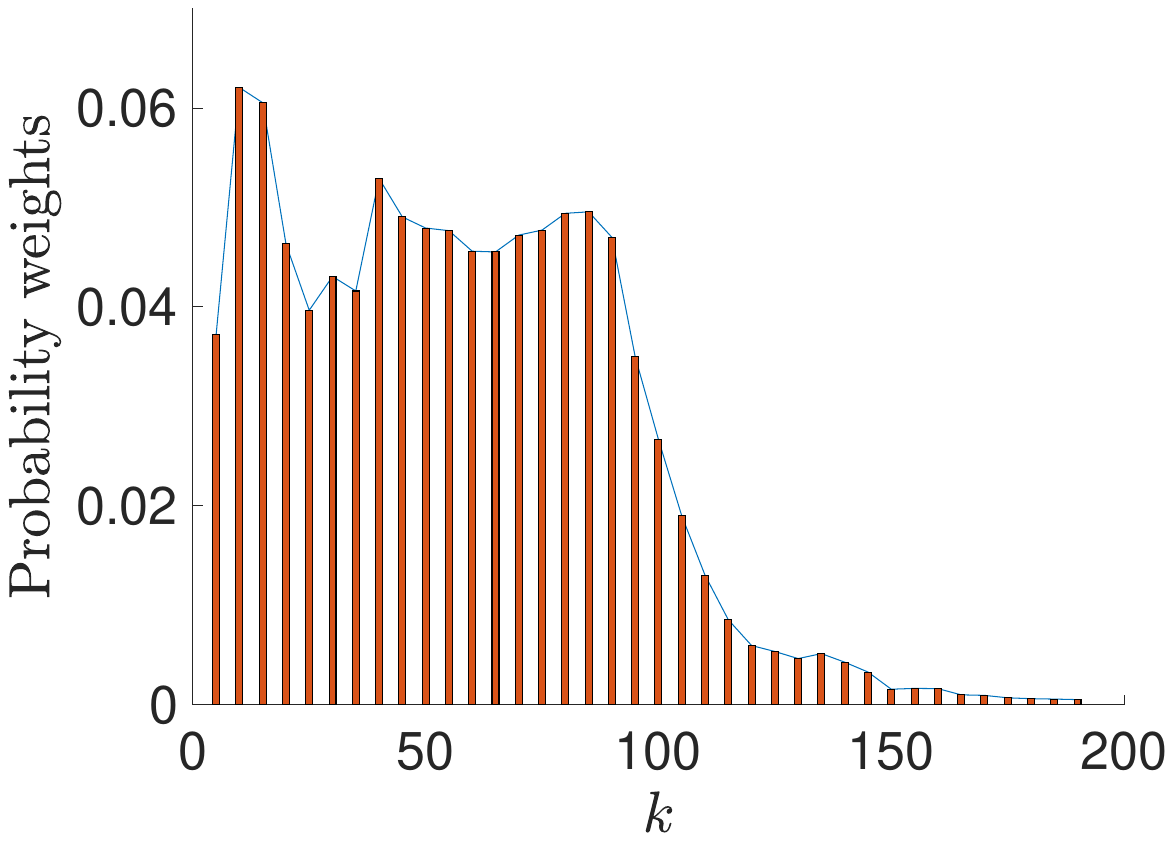}
\caption{Empirical distribution for $k=5,\ldots, 190$.} \label{fig:dsp1}
\end{figure}

To adopt the framework of this paper, we parametrize $\mathcal{X}$ by $k=5,10,\ldots, 190$. Using known geographical distance, for each $k$, we associate it with the $k$-NN graph $G_k$ and obtain the Laplacian $\bL_k$ of $G_k$. For each signal $f$, let $\hat{f}(\bL_k,\cdot)$ be the usual GFT of $f$ w.r.t.\ $\bL_k$. To learn a distribution on $\mathcal{X}$, we define the loss function 
\begin{align}\label{sloss}
	\ell(\bL_k,f)^2 = \frac{\sum_{6\leq i\leq 194}|\hat{f}(\bL_k,i)|^2}{\norm{f}^2}.
\end{align} 
Intuitively, the loss function $\ell$ computes the fraction, in norm, of the high frequency components of $f$ w.r.t.\ $\bL_k$. For sampling, we want to find $G_k$ that best compresses the signals, as we want to sample at only a few stations. Therefore, it is reasonable to choose this loss function.      
To construct the empirical risk with $\ell$ in the Bayesian framework in \cref{sec:lea}, we use $30$ randomly chosen signals, less than $10\%$ of the total number of signals. The resulting learned distribution $\mu_{\mathcal{X}}$ over the parameter space $\set{5,\ldots,190}$ of $\mathcal{X}$ is shown in \figref{fig:dsp1}. We see that local peaks occur at $k=10$ and $k=40$, with the weights dropping sharply after $k=90$. Based on the observation, we further restrict $\mathcal{X} = \{5,\ldots,40\}$.  

For the sampling task, we follow \cref{sec:ban}. We choose $\mathcal{Y}$ to be $\mathcal{X}\times [5]$ and obtain $\bB_{\mathcal{Y}}$ using the empirical distribution $\mu_{\mathcal{X}}$. We apply the sampling and recovery procedure described in \cref{sec:ban} by choosing $V_{189}$ (cf.\ \cref{prop:soo}) consisting of $5$ stations. For each $f$, let $f'$ be the recovered signal as in \cref{prop:soo} using the recovery matrix $\bG_{V_{189}}$. We evaluate the performance by computing \emph{mean error $\mathcal{E}: = \big(\sum_{v\in V}|f(v)-f'(v)|\big)/n$} between $f$ and $f'$ over all stations. On the other hand, for $k=5,\ldots, 90$, we apply the same procedure with the delta distribution $\delta_k$ at $k$ on $\mathcal{X}$. It is nothing but the procedure of recovery of bandlimited signals with $\bL_k$ as in classical GSP. 

\begin{figure}[!htb] 
\centering
\includegraphics[width=0.9\columnwidth,height=.2\textheight,trim=0 0cm 0 0cm, clip]{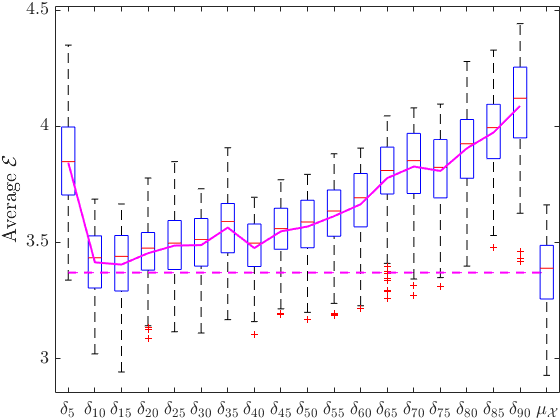}
\caption{Boxplots of average error $\mathcal{E}$. The solid magenta curve shows the mean for each case. It has local minimums as compared to the peaks of \figref{fig:dsp1}. The dashed magenta line shows the mean for $\mu_{\mathcal{X}}$.} \label{fig:dsp14}
\end{figure}

The stations are sampled randomly. However, $\bG_{V_{189}}$ associated with $V_{189}$ can be close to being singular. We perform $200$ trials with non-singular $\bG_{V_{189}}$. For each trial, we compute the average of $\mathcal{E}$ over the whole year. The same is done for $\delta_k, k=5,\ldots,90$. Boxplots of the results are shown in \figref{fig:dsp14}. We see that working with $\mu_{\mathcal{X}}$ has the overall best performance as compared with any $\delta_k$. 

\begin{figure}[!htb] 
\centering
\includegraphics[width=0.65\columnwidth,height=.15\textheight,trim=0 0cm 0 0cm, clip]{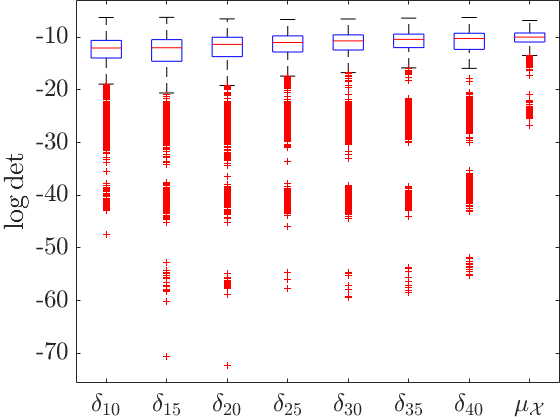}
\caption{Boxplots of $\log\det$ of the recovery matrices.} \label{fig:dsp13}
\end{figure}

For either $\delta_k$ or $\mu_{\mathcal{X}}$ and chosen $V_{189}$, the determinant $\det(\bG_{V_{189}})$ of $\bG_{V_{189}}$ is another indication of sampling quality, as almost singular $\bG_{V_{189}}$ does not permit good recovery (e.g., the default numerical precision of MATLAB is $16$ digits). We randomly sample $V_{189}$ and show, in \figref{fig:dsp13}, boxplots of $\log\det(\bG_{V_{189}})$ for different distributions used. We observe that with $\mu_{\mathcal{X}}$, the recovery matrix is less likely to be singular.

\subsection{ECoG dataset: anomaly detection} \label{sec:ecg}

In this experiment, we perform anomaly detection using band-pass filters (cf.\ \cref{sec:ban}). We consider the ECoG dataset corresponding to two periods (so-called ``pre-ictal'' and ``ictal'') of a seizure in an epilepsy patient.\footnote{\url{https://math.bu.edu/people/kolaczyk/datasets.html}} ECoG signals are measurements taken at each of the $76$ electrodes in the brain of the patient. We test the performance of our framework with the anomaly detection task, by treating ``pre-ictal'' signals as normal and ``ictal'' signals as abnormal. We preprocess each signal by normalizing it to have a unit length. 

For graph construction, we follow \cite{Kra08}. Briefly, there are $76$ nodes associated with $76$ electrodes. To construct a graph, one first computes signal correlations between pairs of nodes. For a chosen $\tau<1$, the graph $G_{\tau}$ is then obtained by thresholding pairwise correlations with $\tau$. We form a sample space of graphs as $\mathcal{X} = \set{G_{\tau} \given \tau=0.25, 0.3, \ldots, 0.8}$. Let $\bL_{\tau}$ be the Laplacian of $G_{\tau}$, and its Fourier basis is $\set{u_{\tau,i}\given i\leq 76}$.

For each $\tau$, by preliminary inspection, we notice that normal (pre-ictal) signals tend to have smaller high frequency components. This prompts us to compute for a signal $f$, the norm $e_{f,\tau}$ of a high-pass filter applied to $f$ as: $e_{f,\tau}^2 = \sum_{i=60}^{76}\langle u_{\tau,i},f\rangle^2$. For each $\tau$, we assume that there is a known $\epsilon_\tau$, and $f$ is declared to be abnormal if $e_{f,\tau}>\epsilon_\tau$. Here, $\epsilon_{\tau}$ is obtained by average $e_{f,\tau}$ for a small sample of both pre-ictal and ictal signals. By going through every $\tau$, we notice that the top $3$ parameters are $\tau = 0.35, 0.4, 0.55$ with accuracy $76.4\%, 76.0\%, 74.6\%$ respectively.  

To apply our framework, we first estimate an empirical distribution of $\mu_{\mathcal{X}}$ following \cref{sec:lea}. We randomly choose a sample consisting of $\kappa$ fraction of all signals. The label $y_f$ for a signal $f$ is $1$ if $f$ is ictal and $0$ otherwise. We modify the $0$-$1$-loss (cf.\ \cite{Gue19} Section 2) for the loss function $\ell(\bL_{\tau},f, y_f) = |1(e_{f,\tau}>\epsilon_{\tau})-y_f|$. The empirical distribution $\mu_{\mathcal{X}}$ depends on both $\kappa$ and chosen samples.

\begin{figure}[!htb] 
\centering
\includegraphics[width=0.6\columnwidth,height=.15\textheight]{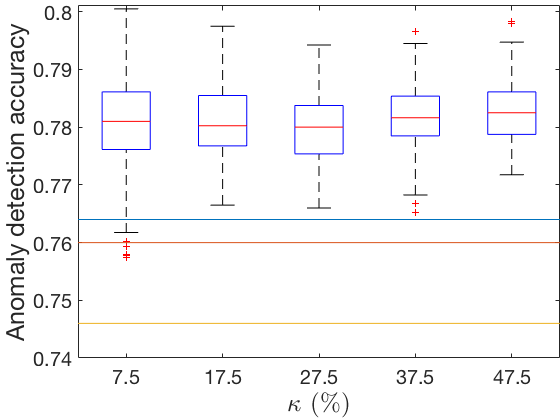}
\caption{The figure shows boxplots of the anomaly detection accuracy. For comparison, horizontal lines show the performance for using single graph operators: $\bL_{0.35}$ (top, blue), $\bL_{0.4}$ (middle, red), and $\bL_{0.55}$ (bottom, yellow).} \label{fig:dsp11}
\end{figure}

For anomaly detection, given signal $f$, we aggregate the normalized difference associated with high-pass filter norms $e_{f,\tau}$ and $\epsilon_{\tau}$: $b_f = \mathbb{E}_{\mu_{\mathcal{X}}} [e_{f,\tau}-\epsilon_{\tau}]/\epsilon_{\tau}$. The signal $f$ is declared to be abnormal if $b_f>0$. We show the detection accuracy in \figref{fig:dsp11} for different $\kappa$. We see that the distributional approach generally outperforms using any single $\bL_{\tau}$. As $\kappa$ increases, the general trend is that the performance improves and the standard deviation decreases. However, both changes are very gradual, and in practice $\kappa \approx 17.5\%$ seems to be sufficient. 

\begin{figure}[!htb] 
\centering
\includegraphics[width=0.6\columnwidth,height=.15\textheight]{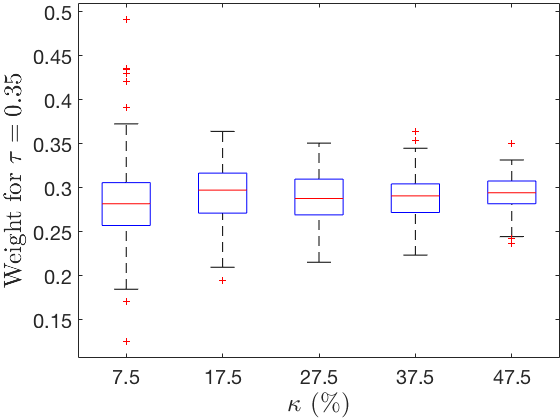}
\caption{Boxplots of the probability weight for $\bL_{\tau}$, where $\tau=0.35$, in $\mu_{\mathcal{X}}$.} \label{fig:dsp12}
\end{figure}

We have noticed that for a single operator, $\bL_{0.35}$ performs the best. We want to investigate its role in the distributional approach by computing its probability weight in each $\mu_{\mathcal{X}}$. The results are shown in \figref{fig:dsp12}. We notice that as $\kappa$ increases, the standard deviation decreases as expected. On the other hand, the median stays approximately constant near $0.3$. This suggests that contributions from operators other than $\bL_{0.35}$ are also significant.

\subsection{Network infection spreading: base change} \label{sec:nis}

We demonstrate base change (cf.\ \cref{sec:ba}) in this subsection. Continuing from \cref{eg:sap}, we assume that transmission between certain vertices in the graph is fast. For example, two co-workers in the same office may receive a piece of information at almost the same time or may display symptoms of a disease at around the same time. The collection of such connections $F$ is a subset of edges. Suppose we are given samples $\scD$ of pairs $(s,\mathcal{T})$, where any sample in $\scD$ does not contain edges with fast transmission. For example, the training data based on contact tracing during a disease lockdown may not have included infections between co-workers while the new observations after lifting the lockdown do, and we need to learn an updated distribution for the current observations. The approach illustrated here can be adapted and applied to other types of changes in the graph attributes. To account for this in our inference, we need to perform a base change. 

\begin{figure}[!htb] 
\centering
\includegraphics[scale=0.35]{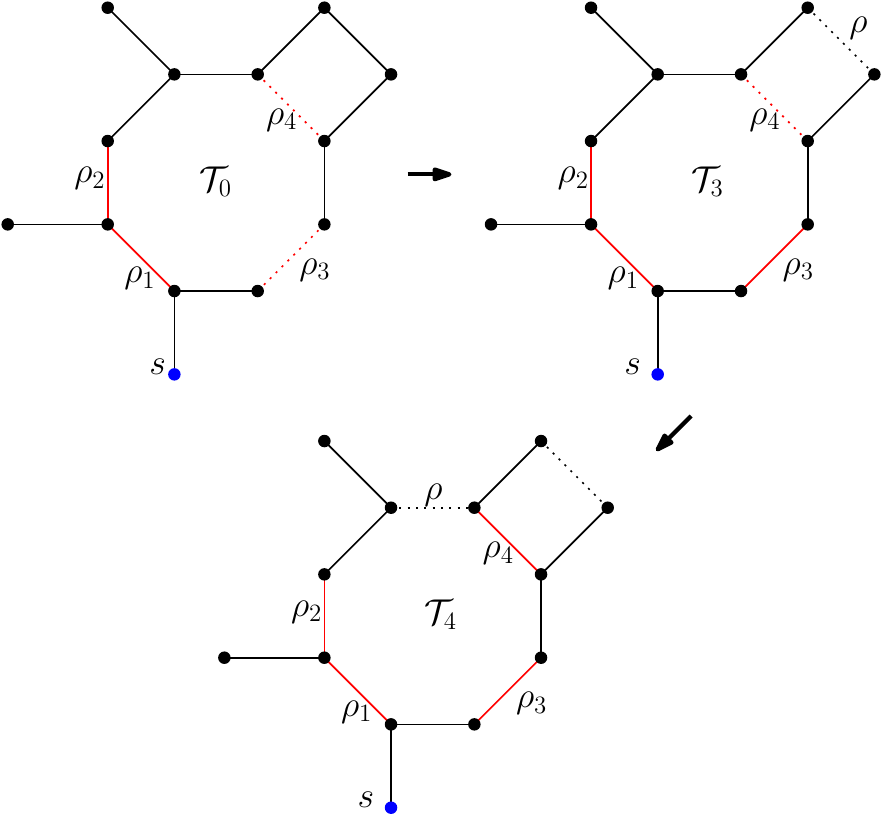}
\caption{We give an example of $h: \calZ \to \mathcal{X}$: the red edges belong to $F=\{\rho_1,\rho_2,\rho_3,\rho_4\}$ and dotted edges are not contained in the spanning tree. The blue node $s$ is the source. Starting from $\mathcal{T}_0$, we have $\mathcal{T}_1=\mathcal{T}_2=\mathcal{T}_0$ (not shown in the illustration) as both $\rho_1,\rho_2$ are contained in $\mathcal{T}_0$. To obtain $\mathcal{T}_3$, we want to include $\rho_3$. We first identify the path in $\mathcal{T}_2=\mathcal{T}_0$ that connects the endpoints of $\rho_3$, which is the big cycle removing $\rho_3$ with $9$ edges. Seven of them do not belong to $F$. Among them, we pick $\rho$ which has the median distance to $s$. The tree $\mathcal{T}_3$ is obtained by adding $\rho_3$ and removing $\rho$. Similarly, from $\mathcal{T}_3$, we add $\rho_4$ and remove the edge next to it to obtain $\mathcal{T}_4$, and $h(s,\mathcal{T}_0)=(s,\mathcal{T}_4)$.} \label{fig:ct}
\end{figure}

For ease of analysis, assume that $F$ does not contain cycles. Let $\calZ$ be the space of adjacency matrices corresponding to propagation paths that may be without edges from $F$ and $\calX$ be the space containing adjacency matrices that have all edges in $F$. We define a map $h: \calZ \to \mathcal{X}$ as follows (an example is given in \figref{fig:ct}): 
\begin{enumerate}[(a)]
\item Given a pair $(s,\mathcal{T})$, let $\{\rho_1,\ldots,\rho_k\}$ be the edges in $F$ sorted according to the edges' distances to $s$ (the smaller of the distances of $s$ to either end points of each edge). At step $i=0$, we let $\mathcal{T}_0=\mathcal{T}$.
\item Suppose in the $(i-1)$-th step, we have a spanning tree $\mathcal{T}_{i-1}$. If $\rho_i \in F$, we do not make changes and set $\mathcal{T}_i=\mathcal{T}_{i-1}$. Otherwise, suppose the endpoints of $\rho_{i}$ are $v_i, w_i$. There is a unique path $P$ in $T_{i-1}$ connecting $v_i$ and $w_i$. We find an edge $\rho$ in $P\backslash F$ whose distance to $s$ is the median among all edges in $P\backslash F$. Let $\mathcal{T}_i = \mathcal{T}_{i-1}\cup \set{\rho_i} \backslash \set{\rho}$.
\end{enumerate}

Suppose that a snapshot observation of the infection status of the graph vertices at a particular time is given. We want to infer the source $s$. The snapshot observation gives rise to a graph signal $f$ which is $1$ at infected nodes and $0$ otherwise. We may interpret the source identification problem as learning a distribution on $\mathcal{X}$ with a single $f$, and marginalize $\mathcal{T}$ to find $s$ with the largest likelihood. For a $(s,\mathcal{T}) \in \mathcal{X}$, we define the loss $\ell$ as follows: Write $A_{\mathcal{T}}$ for the sum of the identity matrix and the adjacency matrix of $\mathcal{T}$. Let $\tau$ be the function that sends any non-zero components of a vector to $1$, and $\delta_s$ be the signal of unit length that is supported at $s$. If $H_{\mathcal{T}}$ is the height of $\mathcal{T}$ rooted at $s$, then we take $\ell((s,\mathcal{T}), f) = \min_{0\leq i\leq H_{\mathcal{T}}}\norm{\tau \big(A_{\mathcal{T}}^i(\delta_s)\big)-f}$, where $A_{\mathcal{T}}^i$ is applying the shift $A_{\calT}$, $i$ times.

Let $\mu_{\calZ}$ be the empirical probability mass function on the training set $\scD$. The distribution $\mu_{\mathcal{X}}$ on $\mathcal{X}$ sought after is proportional to $\exp(-\gamma \ell(\cdot,f))h_*(\mu_{\calZ})$ (cf.\ \cref{sec:ba}).

We perform simulations on a $2$D-lattice and Enron email graph with $225$ and $300$ nodes. The source $s$ is uniformly randomly chosen from a list of $20\%$ candidates, and the propagation path $\mathcal{T}$ is uniformly randomly chosen from the breadth-first search (BFS) trees rooted at $s$, which is a reasonable spreading assumption \cite{Shah2011}. The snapshot observation is made when around $40\%$ of nodes are infected. We run simulations for $F$ with sizes $20\%$, $40\%$, $60\%$, and $80\%$ of $|E|$. If $s^*$ is the estimated source node, we evaluate the performance by measuring its distance $d$ to the actual source $s$. For comparison, we summarize, in \cref{tab:dsp3}, the percentage improvement in the error distance with the base change using the mapping $h$ described above, over that without any base change.

In general, we do benefit from the base change. This is more prominent when $|F|$ is large, as expected. For small $|F|$, the performance working without base change may have a slightly better performance.

\begin{table}[!htb]
\centering
\caption{Improvement in error distance with base change.} \label{tab:dsp3}
\begin{subtable}[t]{0.8\columnwidth}
\centering
\begin{tabular}{|r|c|c|c|c|c|c|c|c|c|} 
\hline
$|F|/|E|$ & $20\%$ & $40\%$ & $60\%$ & $80\%$ \\ 
\hline
\% improvement & $-5.6\%$ & $15.2\%$ & $77.6\%$ & $68.4\%$ \\ 
\hline
\end{tabular}
\caption{$2$D-lattice}
\label{dsp3:2D}
\end{subtable}

\vspace{.1in}

\begin{subtable}[t]{0.8\columnwidth}
\centering
\begin{tabular}{|r|c|c|c|c|c|c|c|c|c|} 
\hline
$|F|/|E|$ & $20\%$ & $40\%$ & $60\%$ & $80\%$ \\ 
\hline
\% improvement & $9.65\%$ & $27.7\%$ & $41.2\%$ & $19.1\%$ \\ 
\hline
\end{tabular}
\caption{Enron email graph}
\label{dsp3:Enron}
\end{subtable}
\end{table}   

\section{Conclusion} \label{sec:con}
We have presented a new GSP framework over a probability space of shift operators. This is useful in applications where the underlying graph topology is uncertain or where we do not know \emph{a priori} what is a shift operator consistent with the observations. We develop the concepts of Fourier transform, MFC filters, and band-pass filters. We discuss and develop methods to allow a change of the underlying probability space of shift operators, which we call a base change. Finally, the usefulness of our framework is demonstrated with numerical experiments on both synthetic and real datasets. 

For future work, we will explore how to gain knowledge of the distribution with even less prior information, so that the current framework can be applied. Complex graph Fourier transform and the more general GGSP have been proposed \cite{Bel19, Ji19}. It may bring new insights by synergizing these frameworks with our probabilistic approach. Another interesting future research direction is to explore the possibility of using the framework to develop new graph neural network methods.   

\appendices

\section{Proofs of theoretical results}\label[Appendix]{sec:pro}

\begin{IEEEproof}[Proof of \cref{lem:Gamma_exp}]
From \cref{def:conv}, it is clear that $\bigstar_{\Gamma}$ is linear. To show that it is bounded, taking the norm of \cref{def:conv}, we have from the triangle and Cauchy-Schwarz inequalities,
\begin{align*}
\norm{\bigstar_{\Gamma}(f)} 
&\leq \int_{\mathcal{X}} \sum_{i=1}^n |\Gamma(\bX,i)| \cdot \norm{f} \ud\mu_{\mathcal{X}}(\bX)  \\
&\leq C \norm{\Gamma}_{L^2(\mathcal{X}\times[n])} \cdot \norm{f},
\end{align*}
for some constant $C$. To show the expectation form, we note that for all $f\in L^2(V)$,
\begin{align*}
\bigstar_{\Gamma}(f)(v) 
&= \int_{\mathcal{X}} \sum_{i=1}^n \Gamma_{\bX}(i)\ip{f}{u_{\bX,i}} u_{\bX,i}(v) \ud\mu_{\mathcal{X}}(\bX)\\
&= \int_{\mathcal{X}} \bigstar_{\Gamma_{\bX}}(f) (v) \ud\mu_{\mathcal{X}}(\bX) = \E_{\mu_{\mathcal{X}}}[\bigstar_{\Gamma_{\bX}}](f)(v),
\end{align*}
where the last equality holds because $\bigstar_{\Gamma_{\bX}}$ is a linear operator and can hence be written as an $n\times n$ matrix whose entries are functions of $\bX$. The result then follows from an interchange of the integral and finite sum. 
\end{IEEEproof}

\begin{IEEEproof}[Proof of \cref{lem:acf}]
If $\bX\in \mathcal{X}$ does not have repeated eigenvalues, then the space of fiberwise convolutions $\bigstar_{\Gamma_{\bX}}$ is isomorphic to the space of degree $n-1$ polynomials in $\bX$ (cf.\ \cite{Shu13}). Moreover, we have seen that $\bigstar_{\Gamma}$ is the expectation of $\bigstar_{\Gamma_{\bX}}$ from \cref{lem:Gamma_exp}, and the result follows. 
\end{IEEEproof}

\begin{IEEEproof}[Proof of \cref{thm:bipoly}]
We remark that the space of MFC filters on $L^2(V)$ is finite-dimensional, as it is a subspace of the finite-dimensional space $M_n(\mathbb{R})$ (see \cref{lem:Gamma_exp}). We want to show that for each $\epsilon>0$ and $\bigstar_{\Gamma}$, there is a bi-polynomial filter $\bF$ such that the $\norm{\bF-\bigstar_{\Gamma}}\leq \epsilon$, where $\norm{\cdot}$ is the operator norm. However, all bi-polynomial filters form a subspace of the space of MFC filters. The above approximation property cannot hold for these two finite-dimensional vector spaces unless they are the same.

Let $c>0$ be the upper bound on the operator norm of $\bX_t^i, t\in T$, $0\leq i\leq n-1$ for almost every $\bX\in \mathcal{X}$. For $\bX_t \in \mathcal{X}$ with no repeated eigenvalues, $\bigstar_{\Gamma_{\bX_t}} = \sum_{0\leq i\leq n-1}a_i(t)\bX_t^i$ for some $a_i(t) \in L^2(T)$. The Weierstrass approximation theorem \cite{Rud76} says that any continuous function on $T$ can be approximated arbitrarily closely by a polynomial on $T$ with the uniform norm. Moreover, the space of continuous functions is dense in $L^2(T)$ \cite{Rud87}. As a consequence, we can find a polynomial $b_i(t)$ such that $\norm{b_i(t)-a_i(t)}_{L^2(T)}$ is as small as we wish, say bounded by $\epsilon/nc$. Let $\bigstar_{\Gamma'_{\bX_t}} = \sum_{0\leq i\leq n-1}b_i(t)\bX_t^i$. We have 
\begin{align*}
& \norm*{\E_{\mu_T}[\bigstar_{\Gamma_{\bX_t}}]-\E_{\mu_T}[\bigstar_{\Gamma'_{\bX_t}}]}_{L^2(T)} \\
= & \norm*{\E_{\mu_T} \sum_{0\leq i\leq n-1}(a_i(t)-b_i(t))\bX_t^i}_{L^2(T)} \\
 \leq & \sum_{0\leq i\leq n-1} \norm{b_i(t)-a_i(t)}_{L^2(T)} \cdot c
\leq n \frac{\epsilon}{nc} c = \epsilon.
\end{align*}
This proves the claim of the first paragraph and hence the theorem.
\end{IEEEproof}

\begin{IEEEproof}[Proof of \cref{lem:tso}]
For $(\mathcal{Y},\epsilon)$-bandlimited signals $f_1,f_2$ and $0\leq a\leq 1$, let $f = af_1+(1-a)f_2$. We have $\norm{\bB_{\mathcal{Y}}(f)-f} \leq a\norm{\bB_{\mathcal{Y}}(f_1)-f_1}+(1-a)\norm{\bB_{\mathcal{Y}}(f_2)-f_2} \leq \epsilon$. This shows that $(\mathcal{Y},\epsilon)$-bandlimited signals form a convex set. If $\bB_{\calY}$ does not fix any 
non-zero signal, i.e., $\bB_{\calY}(f)\neq f, f\neq 0$, then $\bB_{\calY}-\bI$ is invertible. Let $\lambda_{\calY}$ be eigenvalue of $\bB_{\calY}-\bI$ smallest in magnitude. We have $|\lambda_{\calY}|>0$. Hence, $\norm{\bB_{\mathcal{Y}}(f)-f}\geq |\lambda_{\calY}|\norm{f}$. Therefore, if $f$ is $(\calY,\epsilon)$-bandlimited, $\norm{f}\leq \epsilon/|\lambda_{\calY}|$.

Moreover, from \cref{lem:Gamma_exp} and the fact that norms are continuous, $f\mapsto \norm{\bB_{\mathcal{Y}}(f)-f}$ defines a continuous function $\mathbb{R}^n \to \mathbb{R}$. If $\epsilon>0$, the inverse image of $(-\epsilon,\epsilon)$ is an open subset of $\mathbb{R}^n$ containing $0$. Hence, $0$ is an interior point of the set of $(\mathcal{Y},\epsilon)$-bandlimited signals.
\end{IEEEproof}

\begin{IEEEproof}[Proof of \cref{lem:ate}]
From \cref{lem:Gamma_exp}, we have $\bB_{\mathcal{Y}} = \E_{\mu_{\mathcal{X}}}[\bF_{\bX}]$. In the expression, $\bF_{\bX}$ is a graph band-pass filter (in classical GSP \cite{San13} Section V.A.) parametrized by $\bX$ and its eigenvalues belong to $[0,1]$.  We want to show that $\bB_{\mathcal{Y}}$ is positive semi-definite and its operator norm is bounded by $1$.

For any signal $f\in L^2(V)$, we have 
\begin{align*}
\norm{\bB_{\mathcal{Y}}(f)} = \norm*{\E_{\mu_{\mathcal{X}}}[\bF_{\bX}](f)} \leq \E_{\mu_{\mathcal{X}}}\norm{\bF_{\bX}(f)} \leq \norm{f}.
\end{align*}
On the other hand, we also have the lower bound
\begin{align*}
    \ip{\bB_{\mathcal{Y}}(f)}{f} = \int_{\mathcal{X}} \ip{\bF_{\bX}(f)}{f}\ud\mu_{\mathcal{X}}(\bX) \geq \int_{\mathcal{X}} 0 \ud\mu_{\mathcal{X}} = 0.
\end{align*}
\end{IEEEproof}

\begin{IEEEproof}[Proof of \cref{lem:sfi}]
As $f = \sum_{1\leq i\leq n}a_ie_i$, $\bB_{\mathcal{Y}}(f)-f = \sum_{1\leq i\leq n} a_i(\lambda_i-1)e_i$. Using orthogonality of $e_i$, $1\leq i\leq n$, we have $\norm{\bB_{\mathcal{Y}}(f)-f}^2 = \sum_{1\leq i\leq n}(1-\lambda_i)^2a_i^2$. As $f$ is $(\mathcal{Y},\epsilon)$-bandlimited, $\sum_{1\leq i\leq n}(1-\lambda_i)^2a_i^2\leq \epsilon^2$. Therefore, $\sum_{1\leq i\leq j}a_i^2\leq \epsilon^2/(1-\lambda_j)^2.$
\end{IEEEproof}
\vspace{.1in}

\begin{IEEEproof}[Proof of \cref{prop:soo}]
\begin{enumerate}[(a)]
\item Suppose we express $f$ as a vector in the coordinates given by the basis $u_1,\ldots,u_n$. We decompose $f = f_1+f_2$ where $f_1$ belongs to the span of $u_1,\ldots, u_j$ and $f_2$ belongs to the span of $u_{j+1},\ldots, u_n$. Correspondingly, we express $f_{V_j}= f_{1,V_j}+f_{2,V_j}$ where $f_{i,V_j},i=1,2$ takes the $V_j$-components of $f_i$. Hence, $f' = \bU_{>j}\bG_{V_j}^{-1}(f_{1,V_j})+\bU_{>j}\bG_{V_j}^{-1}(f_{2,V_j})$. 

As $\bG_{V_j}$ is the recovery matrix associated with the uniqueness set $V_j$, we have $f_2 = \bU_{>j}\bG_{V_j}^{-1}(f_{2,V_j})$. Therefore, $\norm{f'-f} = \norm{f_1-\bU_{>j}\bG_{V_j}^{-1}(f_{1,V_j})}$. 
On the other hand, since $f$ is $(\mathcal{Y},\epsilon)$-bandlimited, $\norm{f_1} \leq \epsilon/(1-\lambda_j)$ from \cref{lem:sfi}. Hence, we have the following estimation:
\begin{align*}
& \norm{f_1-\bU_{>j}\bG_{V_j}^{-1}(f_{1,V_j})}  \leq \norm{f_1} +  \norm{\bU_{>j}\bG_{V_j}^{-1}(f_{1,V_j})} \\
= & \norm{f_1} +  \norm{\bG_{V_j}^{-1}(f_{1,V_j})}\leq \norm{f_1} + \sigma_{V_j}\norm{f_{1,V_j}} \\
 \leq & \norm{f_1}(1+ \sigma_{V_j}) 
\leq  \epsilon\frac{1+\sigma_{V_j}}{1-\lambda_j}.
\end{align*}

\item Using \ref{it:nff} and the condition that $f$ is $(\mathcal{Y},\epsilon)$-bandlimited
\begin{align*}
& \norm{\bB_{\mathcal{Y}}(f')-f'} & \\
= &\norm{\bB_{\mathcal{Y}}(f'-f) + (f'-f) + (\bB_{\mathcal{Y}}(f)-f)} \\ 
 \leq &\norm{\bB_{\mathcal{Y}}(f'-f)} + \norm{f'-f} + \norm{\bB_{\mathcal{Y}}(f)-f}\\
 \leq & 2\norm{f'-f} + \epsilon 
\leq \epsilon \parens*{1+2\frac{1+\sigma_{V_j}}{1-\lambda_j}}.
\end{align*}
\end{enumerate}
The proof is now complete.
\end{IEEEproof}

\section{Convergence of sets of bandlimited signals} \label[Appendix]{sec:cbs}

In this appendix, we consider the following scenario: if we have a sequence of distributions of shift operators $\mu_m$ that converges to the delta distribution $\delta_{\bX_0}$ for some $\bX_0\in\calX$, then we may construct sequences of sets of bandlimited signals as in \cref{def:bandpass}. We are interested in their limits and how they are related to traditional GSP theory. We first formally introduce some notions of convergence. 

In general, let $(S,d)$ be a metric space that is Radon. The \emph{($2$-)Wasserstein distance} $W(\mu,\nu)$ (cf.\ \cite{Vil09}) between probability measures $\mu$ and $\nu$ with finite second moments is given by
\begin{align*}
    W(\mu,\nu)^2 = \inf_{\xi \in \Xi(\mu,\nu)} \int_{(x,y)\in S\times S}d(x,y)^2\ud\xi, 
\end{align*}
where $\Xi(\mu,\nu)$ is the set of joint distributions on $S\times S$ whose marginals are $\mu$ and $\nu$ respectively. This notion of distance is used to define the convergence of probability measures. As we are primarily interested in convergence to a delta distribution, the following explicit formula is useful:
\begin{align}
    W(\mu,\delta_{x_0})^2 = \int_{x\in S} d(x,x_0)^2\ud\mu.
\end{align}

For our purpose, we endow the space $M_n(\mathbb{R})$ of $n\times n$ real-valued matrices with the metric $d_F(\cdot,\cdot)$ induced by the \emph{Frobenius norm}, i.e., $d_F(\bM,\bN) = \norm{\bM-\bN}_F$, for $\bM, \bN \in M_n(\mathbb{R})$, where $\norm{\cdot}_F$ is the Frobenius norm.

With the setup described above, we can rigorously talk about the convergence of $(\mu_m)_{m\geq1}$ to $\delta_{\bX_0}$, where $(\mu_m)_{m\geq1}$ is a sequence of distributions of shift operators on a sample space $\calX$ and $\delta_{\bX_0}$ is the delta distribution supported on the single operator $\bX_0 \in \calX$. For simplicity, we assume all the operators are positive semi-definite, though the results in this appendix hold without this assumption.

Recall that our goal is to study and compare different sets of bandlimited signals, thus we also need a distance measure for sets. For this, we use the Hausdorff metric (cf.\ \cite{Bri99}). Let $S_1$ and $S_2$ be two subsets of a metric space $(S,d)$. Their \emph{Hausdorff metric} $d_H(S_1,S_2)$ is defined by the following expression
\begin{align*}
\max\{\sup_{s_1\in S_1}\inf_{s_2\in S_2}d(s_1,s_2),\sup_{s_2\in S_2}\inf_{s_1\in S_1}d(s_1,s_2)\}.
\end{align*}
Intuitively, it measures how far any point in $S_1$ is away from $S_2$ and vice versa. 

To describe and prove the main result, we revisit \cref{sec:ban}. Fix a subset $J$ of $[n]$. Let $B_0$ be the vector space of $J$-bandlimited signals w.r.t.\ $\bX_0$, i.e, $f \in B_0$ is in the span of $\set{u_{\bX_0,i} \given i\in J}$. Let $\bB_0$ denote the band-pass filter that is the projection onto $B_0$. 

Suppose $(\mu_m)_{m\geq 1}$ is a sequence of probability distributions on $\calX$. Following \cref{sec:ban}, let $\calY = \calX\times J \subset \calX\times [n]$ and $\bB_{\calY, m}$ be the band-pass filter associated with $\calY$ w.r.t.\ $\mu_m$. For $\epsilon\geq 0$, denote by $B_{m,\epsilon}$ the set of $(\calY,\epsilon)$-bandlimited signals associated with $\bB_{\calY, m}$. 

\begin{Theorem} \label{thm:soi}
Suppose 
$\bX_0$ does not have repeated eigenvalues. If $\mu_m$ converges to $\delta_{\bX_0}$ as $m\to \infty$, then there is a sequence of positive numbers $\epsilon_m \to 0$ such that for any compact convex set $K$ containing a sphere $\calS$ (centered at the origin), $B_{m,\epsilon_m}\cap K$ converges to $B_0\cap K$ in the Hausdorff metric. 
\end{Theorem}

\begin{figure}[!htb] 
\centering
\includegraphics[scale=0.38]{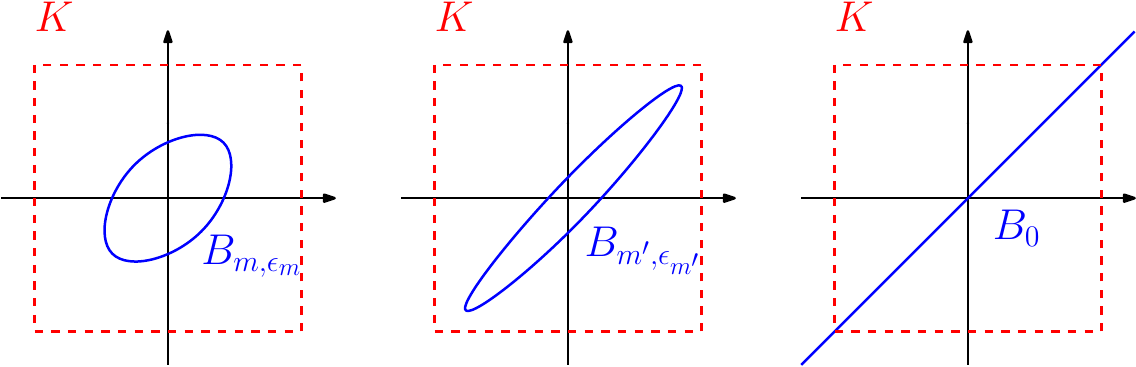}
\caption{From $m$ to $m'>m$, we see that the shape of $B_{m,\epsilon_m}$ bounded by $K$ deforms gradually to that of $B_0\cap K$.} \label{fig:cb}
\end{figure}

Intuitively, the result states that the ``shape'' of $B_{m,\epsilon_m}$ converges to that of $B_0$ (\figref{fig:cb}). Taking intersection with $K$ is necessary, for otherwise, $d_H(B_{m,\epsilon_m}, B_0)$ is $\infty$ for any $m$ with $B_{m,\epsilon_m}$ bounded (cf.\ \cref{lem:tso}),  while $B_0$ is always unbounded.   

We prove \cref{thm:soi} in a few steps. Recall for $\bX \in \calX$, $\lambda_{\bX} = (\lambda_{\bX,i})_{i\in [n]}$ is its ordered set of eigenvalues and $\set{u_{\bX,i} \given i\in [n]}$ are the associated unit eigenvectors. 

\begin{Lemma} \label{lem:fvt}
 For $\varepsilon>0$, there is $\iota>0$ such that if $d_{F}(\bX,\bX_0)\leq \iota$, then $\norm{u_{\bX,i}-u_{\bX_0,i}}\leq \varepsilon$ and $|\lambda_{\bX,i}-\lambda_{\bX_0,i}|\leq \varepsilon$ for all $i\in [n]$.
\end{Lemma}

\begin{proof}
We define 
\begin{align*}
E: \mathbb{R}^n \to \mathbb{R}^n, x= (x_1\ldots,x_n) \mapsto \big((-1)^{i}e_i(x)\big)_{1\leq i\leq n},
\end{align*}
where $e_i$ is the $i$-th elementary symmetric function. Therefore, the components of $E(\lambda_{\bX})$ are the coefficients of the characteristic polynomial $P_{\bX}(t)$ of $\bX$.  

The Jacobian $J(E)$ of $E$ satisfies 
\begin{align*}
|J(E)(x)| = \abs*{\prod_{1\leq i\neq j\leq n}(x_i-x_j)}.
\end{align*}
To see this, we first notice that both sides are polynomials of the same degree. Moreover, if $x_i=x_j, i\neq j$, then $\partial E/\partial x$ has two identical columns and hence both $|J(E)(x)|$ and $|\prod_{1\leq i\neq j\leq n}(x_i-x_j)|$ are $0$. 

As a consequence, $|J(E)(\lambda_{\bX_0})|\neq 0$ and by the inverse function theorem, there is an open neighborhood $U_{\bX_0}$ of $\lambda_{\bX_0}$ that is diffeomorphic to its image $E(U_{\bX_0})$, an open neighborhood of $E(\lambda_{\bX_0})$. As $\lambda_{\bX_0,1}<\ldots<\lambda_{\bX_0,n}$, we may assume that for $x\in U_{\bX_0}$, the entries of $x=(x_i)_{1\leq i\leq n}$ satisfies $x_1<\ldots<x_n$.  

If $d_F(\bX,\bX_0)$ is small enough (i.e., entries of $\bX$ and $\bX_0$ are close), then coefficients of the characteristic polynomial $P_{\bX}(t)$ form a vector in $E(U_{\bX_0})$. Therefore, the vector of eigenvalues $\lambda_{\bX}$ of $\bX$ is in $U_{\bX_0}$ and $\norm{\lambda_{\bX_0}-\lambda_{\bX}}$ can by made arbitrarily small by reducing $d_F(\bX,\bX_0)$. In other words, there is $\iota$ such that if $d_F(\bX,\bX_0)\leq \iota$, then $|\lambda_{\bX,i}-\lambda_{\bX_0,i}|\leq \varepsilon, i\in [n]$. 

To obtain $u_{\bX,i}$, it is one of the intersections of the unit sphere in $\mathbb{R}^n$ with the line $\calL_{\bX,i}$ defined by the equation $(\bX- \lambda_{\bX,i}\bI)x=0$. The lines $\calL_{\bX,i}, \calL_{\bX_0,i}$ can be arbitrarily close to each other if $d_F(\bX,\bX_0)$ (and hence $|\lambda_{\bX,i}-\lambda_{\bX_0,i}|$) is small enough, say $d_F(\bX,\bX_0)\leq \iota$ for small $\iota$. Therefore, $u_{\bX,i}$ can be chosen (as one of the two intersections of $\calL_{\bX,i}$ and the unit sphere) such that $\norm{u_{\bX,i}-u_{\bX_0,i}}\leq \varepsilon$.
\end{proof}

\begin{Lemma} \label{lem:bct}
$\lim_{m\to \infty}\bB_{\calY, m} = \bB_0$ in operator norm. 
\end{Lemma}

\begin{proof}
Let $\norm{\cdot}_{\text{op}}$ be the operator norm. It is a general fact on finite dimensional spaces that it is equivalent to the Frobenius norm, i.e., the notion of convergence is the same in both norms.  

For $\bX \in \calX$, let $\bP_{J,\bX}$ be the projection matrix to the space spanned by $\{u_{\bX,i},i\in J\}$. By \cref{lem:fvt}, the operator norm of $\bP_{J,\bX}$ can be arbitrarily close to that of $\bP_{J,\bX_0}$ if $d_F(\bX,\bX_0) = \norm{\bX-\bX_0}_F$ is small enough. This means for $\varepsilon>0$, there is $\iota>0$ such that $\norm{\bX-\bX_0}_F\leq \iota$ implies that $\norm{\bP_{J,\bX}-\bP_{J,\bX_0}}_{\text{op}}\leq \varepsilon$. 

On the other hand, the operator $\bB_{\calY, m}$ is defined by $\bB_{\calY, m}(f) = \mathbb{E}_{\bX\sim \mu_m}\bP_{J,\bX}(f), f\in \mathbb{R}^n$. Consider any measurable subset $U_m$ of $\calX$ and its complement $U_m^c$. For any unit vector $f$, we estimate
\begin{align*}
    \norm{\bB_{\calY, m}(f) - \bB_0(f)} 
    & \leq \int_{\bX\in U_m}\norm{\bP_{J,\bX}(f)-\bP_{J,\bX_0}(f)}\ud \mu_m \\
    & + \int_{\bX\in U_m^c}\norm{\bP_{J,\bX}(f)-\bP_{J,\bX_0}(f)}\ud \mu_m \\
    \leq & 2\mu_m(U_m) + \sup_{\bX\in U_m^c}\norm{\bP_{J,\bX}-\bP_{J,\bX_0}}_{\text{op}}. 
\end{align*}
Recall that the Wasserstein metric $W(\mu,\delta_{\bX_0})$ satisfies, $W(\mu,\delta_{\bX_0})^2 = \mathbb{E}_{\bX\sim \mu_m}\norm{\bX-\bX_0}_F^2$. By the Markov inequality, for $a>0$, we have 
\begin{align*}
\mu_m(\norm{\bX-\bX_0}_F\geq \sqrt{a}) \leq \frac{W(\mu_m,\delta_{\bX_0})^2}{a}.
\end{align*}

We choose $U_m = \{\bX\in \calX\mid \norm{\bX-\bX_0}_F\geq \iota\}$, where $\iota$ chosen as in the second paragraph such that $\sup_{\bX\in U_m^c}\norm{\bP_{J,\bX}-\bP_{J,\bX_0}}_{\text{op}}\leq \varepsilon$. Then for any $m$ large enough such that $W(\mu_m,\delta_{\bX_0})^2 \leq \varepsilon\iota^2$, we have $\mu_m(U_m) \leq \varepsilon$. Therefore, $\norm{\bB_{\calY, m}(f) - \bB_0(f)} \leq 3\varepsilon$ and this shows that $\bB_{\calY, m} \to \bB_0$ as $m\to \infty$. 
\end{proof}

\begin{Lemma} \label{lem:lkb}
    Let $K$ be a compact convex set containing a sphere $\calS$ and $\partial K$ be the boundary of $K$. Define $l: \calS \to \partial K$ as follows. For $s \in \calS$, $l(s)$ is the intersection of $\partial K$ and the ray connecting $0$ and $s$. Then $l$ is continuous.
\end{Lemma}

\begin{proof}
 For any $s$, let $(s_i)_{i\geq 1}$ be any sequence of points on $\calS$ that converges to $s$. It suffices to show $l(s_i) \to l(s)$ as $i\to \infty$. Suppose on the contrary that this does not hold. Then there is an $\varepsilon>0$ and a subsequence $(s_{k_i})_{i\geq 1}$ of $(s_i)_{i\geq 1}$ such that $\norm{l(s_{k_i})-l(s)}\geq \varepsilon$. As $\partial K$ is compact, replacing $l(s_{k_i})_{i\geq 1}$ by a subsequence if necessary, we assume that $l(s_{k_i}) \to l(s'), i\to \infty$ for $s'\neq s$. As $s_{k_i}$ is the projection of $l(s_{k_i})$ to $\calS$, we have $s_{k_i}$ converges to $s'$, which is a contradiction.  
\end{proof}

We are now ready to prove \cref{thm:soi}. As $K$ is bounded, we assume that the norm of each $f\in K$ is bounded by $r>0$. For each $m > 0$, let $\alpha_m$ be the operator norm $\norm{\bB_{\calY, m}-\bB_0}_{\text{op}}$ and $\epsilon_m = \sqrt{\alpha_m}$. By \cref{lem:bct}, $\alpha_m \to 0$ as $m\to \infty$. Given $f \in B_0\cap K$, we have 
\begin{align*}
    \norm{\bB_{\calY, m}(f)-f} = \norm{\bB_{\calY, m}(f)-\bB_0(f)}\leq \alpha_m\norm{f}\leq r\alpha_m.
\end{align*}
Therefore, $f$ belongs to $B_{m,\epsilon_m}$ as long as $m$ is large enough such that $r\leq 1/\sqrt{\alpha_m}$.  

On the other hand, for $f\in B_{m,\epsilon_m}$, consider $\tilde{f}= \bB_0(f) \in B_0$. We estimate
\begin{align*}
     \norm{\tilde{f}-f} 
    \leq &\norm{\bB_0(f)-\bB_{\calY, m}(f)} + \norm{\bB_{\calY, m}(f)-f} \\
    \leq & r\alpha_m + \epsilon_m = r\alpha_m + \sqrt{\alpha_m}.
\end{align*}
If $\tilde{f} \in K$, we have $\inf_{f'\in B_0\cap K}\norm{f-f'} \leq r\alpha_m + \sqrt{\alpha_m}$, which converges to $0$ if $m \to \infty$. 

\begin{figure}[!htb] 
\centering
\includegraphics[scale=0.4]{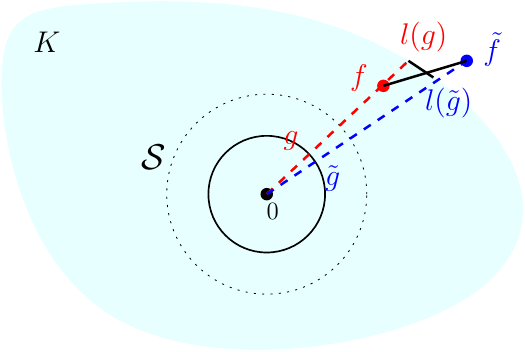}
\caption{Geometric illustration of the proof.} \label{fig:conv}
\end{figure}

Let $r'$ be the radius of $\calS \subset K$. We may assume $m$ is large enough such that if  $\tilde{f} \notin K$, then $\norm{f} \geq r'/2$. Let $g$ and $\tilde{g}$ be the projections of $f$ and $\tilde{f}$ to the sphere of radius $r'/2$ respectively. Hence, the inequality $\norm{g-\tilde{g}} \leq \norm{f-\tilde{f}}$ holds. As in \cref{lem:lkb}, we have $l(g) \in \partial K$ and $l(\tilde{g}) \in \partial K\cap B_0$ (c.f.\ \figref{fig:conv}). By \cref{lem:lkb}, we may assume that $\norm{l(g)-l(\tilde{g})}$ is bounded by $\beta_m$, which converges to $0$ as $m\to \infty$. 

As $f,\tilde{f},l(g),l(\tilde{g})$ are in the same plane (spanned by $f,\tilde{f}$), we apply the triangle inequality to obtain
\begin{align*}
    &\norm{f-l(\tilde{g})} 
     \leq  \norm{f-l(g)} + \norm{l(g)-l(\tilde{g})} \\
    \leq & \norm{f-\tilde{f}} + 2\norm{l(g)-l(\tilde{g})} 
     \leq  r\alpha_m+\sqrt{\alpha_m} + 2\beta_m. 
\end{align*}
Therefore, if $\tilde{f} \notin K$, we have 
\begin{align*}
\inf_{f'\in B_0\cap K}\norm{f-f'} \leq r\alpha_m+\sqrt{\alpha_m} + 2\beta_m,
\end{align*}
which again converges to $0$ if $m \to \infty$.
This concludes the proof that $B_{m,\epsilon_m}\cap K\to B_0\cap K$ in the Hausdorff metric. 

\section{Remarks on algebraic signal processing theory} \label{sec:rmk}

As we have seen, the development of our framework is modeled on the classical GSP theory, though there are many differences. On the other hand, the theory of algebraic signal processing (ASP) \cite{Pus08} is an abstraction of signal processing by summarizing concrete concepts using algebraic languages. In this appendix, we compare with ASP (see also \cite{Jif23} Section V), highlighting their relations and what we need to modify to have a theory that incorporates both probability and algebra. 

Before proceeding further, it is worth mentioning that the fundamental principle we are following is: \emph{The concepts in our framework agree with those in the classical theory if a delta distribution is considered.} There are three concepts we are primarily interested in: the Fourier transform, convolutions, and bandlimited spaces.

Recall ASP requires the data $(\calA, \calM, \rho)$, where $\calA$ is an unital ring (over a base field), $\calM$ is a vector space, and $\rho: \calA \to \text{End}(\calM)$ is an algebra homomorphism of $\calA$ into the endomorphism ring of $\calM$. The map $\rho$ makes $\calM$ an $\calA$-module. Many frameworks do not explicitly refer to ASP. It is because $\calA$ is usually chosen as the polynomial subalgebra of $\text{End}(\calM)$ generated by a single shift operator and $\rho$ is just the inclusion.

In ASP, the Fourier transform is a decomposition of $\calM$ into a direct sum of irreducible submodules $\Delta: \calM \to \oplus_{w\in W}\calM_w$, where $W$ is the index set that also corresponds to the coordinates of the frequency domain. However, in general, if $\calA$ is large, then $\calM$ itself can be irreducible, which makes the setup less useful. In our case, it can be less fruitful by using all the possible shift operators to generate an algebra of endomorphisms. Moreover, we want to encode probabilistic information. Therefore, we extend the codomain of the Fourier transform to a Hilbert space $\calH$, and consider the (Hilbert) homomorphism space $\text{Hom}(\calM, \calH)$, which is no longer a ring. However, there are the fiberwise projections of $\calH \to \calM$. Each composes with a homomorphism in $\text{Hom}(\calM, \calH)$ to give an endomorphism in $\text{End}(\calM)$. Therefore, as we have seen in the paper, the Fourier transform defined in the paper is essentially a fiberwise decomposition into irreducible spaces. 

Though the framework is not convolutional in the sense of \cite{Pus08}, the notion of MFC filters is introduced as an analogy of convolutions in classical GSP for its practical importance. Classically, convolutions allow us to analyze signal interactions in the frequency domain. They are translated into useful operators in the vertex domain via vertex-frequency duality. In theory, there are different perspectives on ``convolution''. In ASP, convolutions are $\rho(\calA) \subset \text{End}(\calM)$. This corresponds to the property that convolution is a polynomial in the generators of $\calA$, as in classical GSP. It is also equivalent to the notion that a convolution corresponds to multiplication by a function in the frequency domain. We take the latter perspective, which leads to \cref{def:convolution}. On the other hand, the polynomial perspective takes its form as in \cref{thm:bipoly}. 

In ASP, a bandlimited space is a submodule of $\calM$, which is isomorphic to the direct sum of irreducible ones. Equivalently, it also corresponds to the invariant subspace of a convolution associated with a characteristic function in the frequency domain. The convolution, called a band-pass filter, is essentially a projection. We take the analogy by introducing a ``band-pass filter'' (cf.\ \cref{def:bandpass}) in the exact same way using MFC filters. However, due to the existence of multiple shift operators with different eigenspaces, there is little hope that a typical band-pass filter has non-trivial invariant spaces. However, we adopt the notion of the space of ``almost invariant vectors'' \cite{Kaz67} in analysis as an alternative (cf.\ \cref{def:bandpass}). It is shown to converge to its counterpart in ASP if the distribution converges to a delta distribution (cf.\ \cref{thm:soi}). In practice, signals hardly belong exactly to any nontrivial submodule perceived in ASP. Approximations are often required for real tasks, and hence we do not see the compromise using spaces of almost invariant vectors as a major setback. 

\bibliographystyle{IEEEtran}
\bibliography{IEEEabrv,StringDefinitions,allref}


\end{document}